\PassOptionsToPackage{dvipsnames}{xcolor}
\documentclass[sigconf, nonacm]{acmart}
\usepackage[utf8]{inputenc}
\usepackage{graphicx} 
\usepackage{url}
\usepackage{amsthm}
\usepackage{pifont}
\usepackage{booktabs}
\usepackage{tabularx}
\usepackage{colortbl}
\usepackage{tikz}
\usetikzlibrary{shapes, arrows.meta, positioning}
\usepackage{soul}
\usepackage{subcaption}
\usepackage{fancyvrb}
\usepackage{float}
\usepackage[subtle]{savetrees}

\newcommand{\hlc}[2][yellow]{{%
    \colorlet{foo}{#1}%
    \sethlcolor{foo}\hl{#2}}%
}

\definecolor{lightgreen}{HTML}{009B55}

\usepackage{fancyhdr}
\fancypagestyle{firststyle}{
   \fancyhf{}
   \fancyhead[C]{This is a preprint}
}

\author{Moe Kayali}
\email{kayali@cs.washington.edu}
\orcid{0000-0002-0643-6468}
\affiliation{University of Washington
\country{}}
\author{Anton Lykov}
\affiliation{University of Washington
\country{}}
\author{Ilias Fountalis}
\affiliation{Relational\textsc{ai}
\country{}}
\author{Nikolaos Vasiloglou}
\affiliation{Relational\textsc{ai}
\country{}}
\author{Dan Olteanu}
\orcid{0000-0002-4682-7068}
\affiliation{University of Zurich
\country{}}
\author{Dan Suciu}
\orcid{0000-0002-4144-0868}
\affiliation{University of Washington
\country{}}

\title{\textsc{Chorus}: Foundation~Models for Unified~Data~Discovery~and~Exploration}
\date{February 2023}

\theoremstyle{definition}
\newtheorem{definition}{Definition}[section]

\begin{abstract}
    We apply foundation models to data discovery and exploration tasks. Foundation models are large language models (\textsc{llm}s) that show promising performance on a range of diverse tasks unrelated to their training. We show that these models are highly applicable to the data discovery and data exploration domain. When carefully used, they have superior capability on three representative tasks: table-class detection, column-type annotation and join-column prediction. On all three tasks, we show that a foundation-model-based approach outperforms the task-specific models and so the state of the art. Further, our approach often surpasses human-expert task performance. We investigate the fundamental characteristics of this approach including generalizability to several foundation models and the impact of non-determinism on the outputs. All in all, this suggests a future direction in which disparate data management tasks can be unified under foundation models.

\end{abstract}

\begin{document}

\newcommand{\rone}[1]{#1}
\newcommand{\rtwo}[1]{#1}
\newcommand{\rthree}[1]{#1}

\maketitle

\section{Introduction}
\label{sec:introduction}

\begin{figure}
    \centering
    \includegraphics[width=0.85\linewidth]{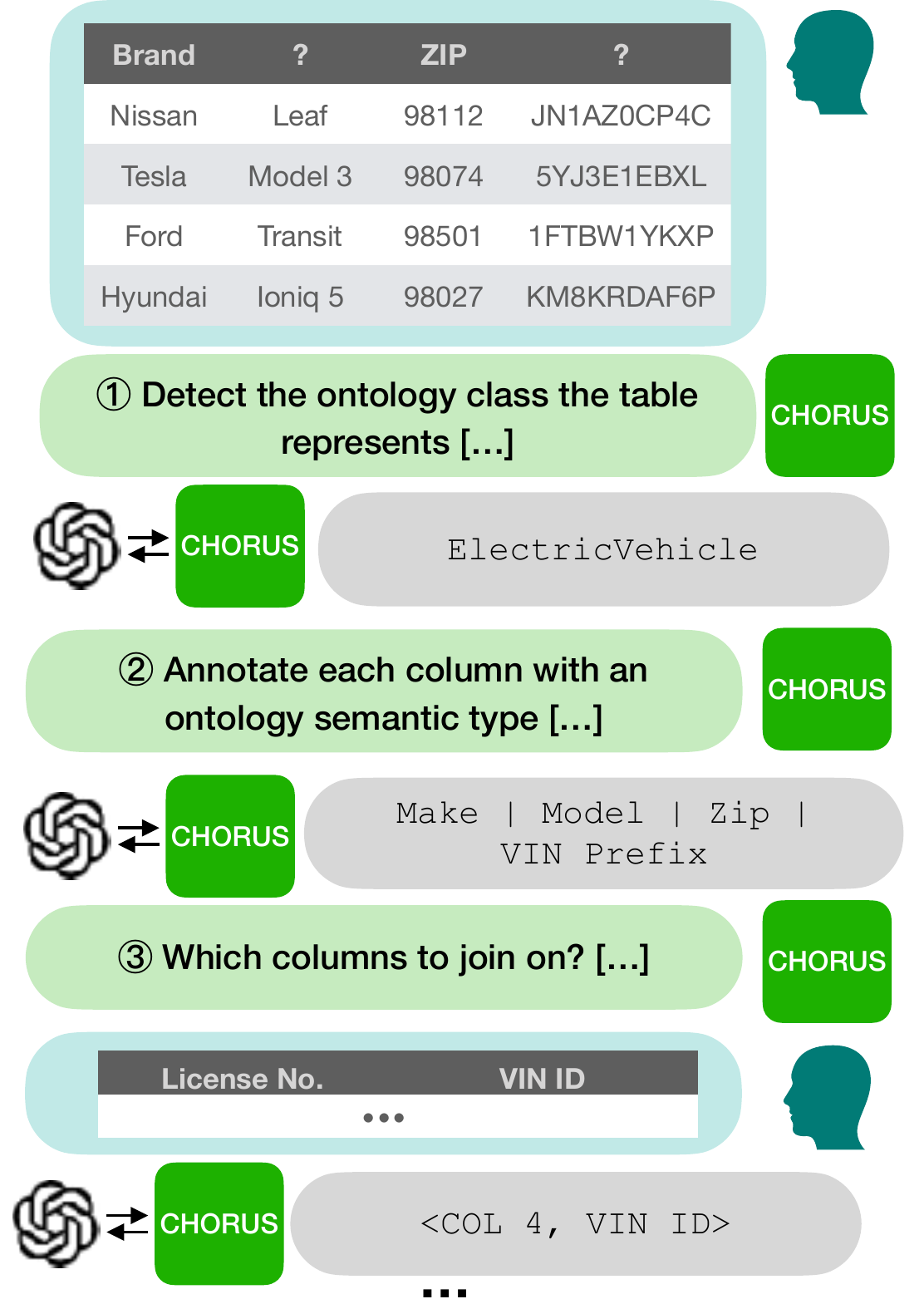}
    \caption{Data discovery tasks considered in this work. Given an ontology, such as DBPedia, \textcircled{1} we assign an overall type to the table and \textcircled{2} we annotate the columns with semantic types. Last, given another table, \textcircled{3} we predict the join column. The user provides the data while \textsc{chorus} interacts with the foundation model. Data from \cite{Licensing:2023aa}, full prompts in Figure~\ref{fig:prompts}.}
    \label{fig:lead_examples}
    \vspace{-0.25in}
\end{figure}

Data discovery and exploration are major components of the workflow of analysts and data scientists. A survey conducted by the Anaconda data-science platform in 2021 found that analysts spend 40\% of their working hours on data loading and cleaning~\cite{Anaconda:2021aa}. Even with this colossal effort, 60-70\% of data within an enterprise still goes unused for analytics~\cite{Gualtieri:2016aa}, remaining as \textit{dark data}~\cite{Heidorn:2008aa,Zhang:2016aa}.

Recent developments in large language-models (\textsc{llm}s) have unlocked human-level performance on diverse domain tasks. The discovery that these models can generalize to diverse domain-specific tasks that they have not been trained on~\cite{Vemprala:2023aa,Wei:2022aa,Andreas:2022aa,Huang:2022aa} has led to emergence of the term \textit{foundation models}~\cite{DBLP:journals/corr/abs-2108-07258}.

Despite their promise, serious risks have hampered the reception of foundation models. These include: spurious generation (including ``hallucination'')~\cite{Holtzman:2020aa}, factual recall limitations~\cite{Mallen:2022aa}, bias~\cite{Gehman:2020aa}, dataset contamination~\cite{Dodge:2021aa}, logical shortcuts~\cite{Shi:2023aa} and fallacies~\cite{Lin:2022aa}. Naïve deployment can lead to unanticipated problems: it has already led to legal action~\cite{Court:2023aa} and recalls by major corporations~\cite{Heaven:2022aa}. These risks are now acknowledged by the creators of these models~\cite{bubeck2023sparks, DBLP:journals/corr/abs-2203-02155, Touvron:2023aa}.

The goal of this paper is to demonstrate the utility of foundation models to the data discovery and exploration while mitigating the aforementioned risks. We select three representative tasks to show the promise of foundation models:  \textcircled{1} \textit{table-class detection}, \textcircled{2} \textit{column-type annotation} and \textcircled{3} \textit{join-column prediction}. An outline of our approach is shown in Figure~\ref{fig:lead_examples}. We call our approach \textsc{chorus}. Code for the system and experiments is available at \url{https://github.com/mkyl/CHORUS}.

\paragraph{Contributions} We summarize our contributions:
\begin{itemize}
    \item[--] The first work to use foundation models for the data discovery tasks of table-class detection, column-type annotation and join-column prediction;
    \item[--] Propose a novel system, \textsc{chorus}, whose flexible architecture enables the synthesis of multiple data discovery tasks and deploying risk mitigations;
    \item[--] Design task-specific approaches that exploit zero- and few-shot strategies and allow information flow between tasks;
    \item[--] Introduce the novel mitigation of \textit{anchoring} to reduce foundation-model risks specific to this domain;
    \item[--] Empirically validate \textsc{chorus}, comparing its performance with the state-of-the-art baselines across three individual tasks.
\end{itemize}

\paragraph{Discussion} Prior work has addressed these tasks individually. Landmark approaches like Sherlock~\cite{DBLP:conf/kdd/HulsebosHBZSKDH19} trained deep model architectures for a specific task, requiring 100K-1M labeled data points. More recent work such as DoDuo~\cite{Suhara:2022aa} and TaBERT~\cite{DBLP:conf/acl/YinNYR20} has focused on \textit{representation learning}, learning embeddings for structured data by improving their performance on one or more downstream tasks.

Foundation models allow a substantially different approach: rather than the classical architecture where the outputs of the model are task-specific, the inputs and outputs of the model are natural language text. Training occurs not on tables or data management tasks specifically, but on general text. Performance on domain-specific tasks is solely by generalization.

This results in a high degree of flexibility. Novel tasks can be specified in natural text, without need for expensive data collection---task examples, metadata and constraints are all incorporated into the task easily. Another advantage of this approach is a \textbf{unified architecture}: tasks can utilize the overall context and previous outputs. For example, in Figure~\ref{fig:lead_examples} the table class \texttt{ElectricVehicle} helps with deducing the outputs \texttt{Make}, \texttt{Model} in the next task.

\paragraph{Outline} Section \ref{sec:background} defines the three tasks investigated in this paper. Section~\ref{sec:approach} describes the architecture of \textsc{chorus} and key approaches. We evaluate the performance of \textsc{chorus} in Section~\ref{sec:experiments}'s experiments. In that section, we also investigate the fundamental characteristics of this approach. We offer a discussion of those results in Section~\ref{sec:discussion}.  This includes a discussion of promising future directions. Finally, we place this work within the literature in Section~\ref{sec:related}, discussing related works. 

\section{Background}
\label{sec:background}
\subsection{Tasks}
\label{sec:tasks}

We assume to be given a {\em data collection} consisting of a number of relational tables $T_1, T_2, \ldots$. Each table $T_i$ consists of a number of columns, or attributes, $A_1, A_2, \ldots$ and a number of rows, or tuples, $r_1, r_2, \ldots$ The name of a table $T_i$ is, in general, non-informative, for example it may be simply a sequential \textsc{id}. The columns may optionally have a name $H_1, H_2, \ldots$ or consist only of values.

In addition to the data collection, we are also given a reference ontology of table classes $C_1, C_2, \ldots$, and a reference ontology of column types $\tau_1, \tau_2,\ldots$. For example, the DBPedia.org types for the table classes include \url{https://dbpedia.org/ontology/Actor} and \url{https://dbpedia.org/ontology/Continent} and column types include \url{https://dbpedia.org/ontology/areaTotal} and \url{https://dbpedia.org/ontology/birthDate}.

We consider three tasks of interest on the data collection:
\begin{definition}[\textcircled{1} Table-class detection]\label{def:table:class:detection}
For each table $T_i$, determine its appropriate class $C_j$, such that every row $r_1, r_2, \ldots$ represents an instance of the $C_j$ type. We adopt this definition from \cite{DBLP:journals/corr/abs-2110-15132}.
\end{definition}

For example, table-class detection on the table given in Figure~\ref{fig:lead_examples} could output \texttt{ElectricVehicle}, since each row of that table is an instance of that class. Alternatively stated, the table is about \texttt{ElectricVehicle}s.

\begin{definition}[\textcircled{2} Column-type annotation] \label{def:column:type:annotation}
For each table $T_i$, find a mapping from its attributes (columns)
  $A_1, A_2, \ldots$ to the reference column types
  $\tau_1, \tau_2, \ldots$, such that each value in $A_i$ is an instance of the $\tau_i$ type. See \cite{DBLP:journals/pvldb/DengSL0020, DBLP:conf/semweb/AbdelmageedCCEH22}.
\end{definition}

For example, column-type annotation on the first column in Figure~\ref{fig:lead_examples} could output \texttt{Manufacturer}, since the values are the respective manufacturers of each \texttt{ElectricVehicle}.

\begin{definition}[\textcircled{3} Join-column prediction]\label{def:join:column:prediction}
    Assume an \textit{execution log} $L$, a history of user actions including table joins and their join conditions, which maps many $(T_i, T_j) \rightarrow (A_k, A_l)$ where $A_k \in T_i, A_l \in T_j$. Given two tables $T$ and $T'$, with columns $A_1, \ldots$ and $A'_1, \ldots$ respectively, the \textit{join-column prediction} task is to suggest a pair $(A_k, A'_l)$ of columns such that the equality condition $A_k=A'_l$, which can be used to join the the tables, matches with the choice in the execution log $L$. For more discussion, see \cite{DBLP:conf/sigmod/YanH20}.
\end{definition}

For example, given the table in Figure~\ref{fig:lead_examples} and another table \texttt{car\_ registration(name, vehicle\_id\_number)},  join-column prediction could output \texttt{(VIN\_prefix, vehicle\_id\_number)}. The correctness of the prediction depends on the ground truth of which columns the user did in-fact join on.

\textbf{Ontologies} Foundation models contain knowledge of ontologies such as DPBedia.org, Freebase and Wikidata. We focus on universal ontologies, that is, ontologies that aim to represent all entities in general. This is in-line with findings that foundation models encode highly technical knowledge, such as clinical reasoning~\cite{Singhal:2022aa} or electrical engineering principles~\cite{Taylor:2022aa}.

\section{Approach}
\label{sec:approach}

We outline the structure of \textsc{chorus} in this section. First, we explore the core idea of ingesting relational data with foundation models and performing data exploration tasks in Subsection~\ref{sec:approach:input}. Next, we describe the necessary post-processing and mitigations we develop in Subsection~\ref{sec:approach:post}.

\begin{figure}
    \centering
    \includegraphics[trim={0 0.9in 0 0}, clip, width=0.9\linewidth]{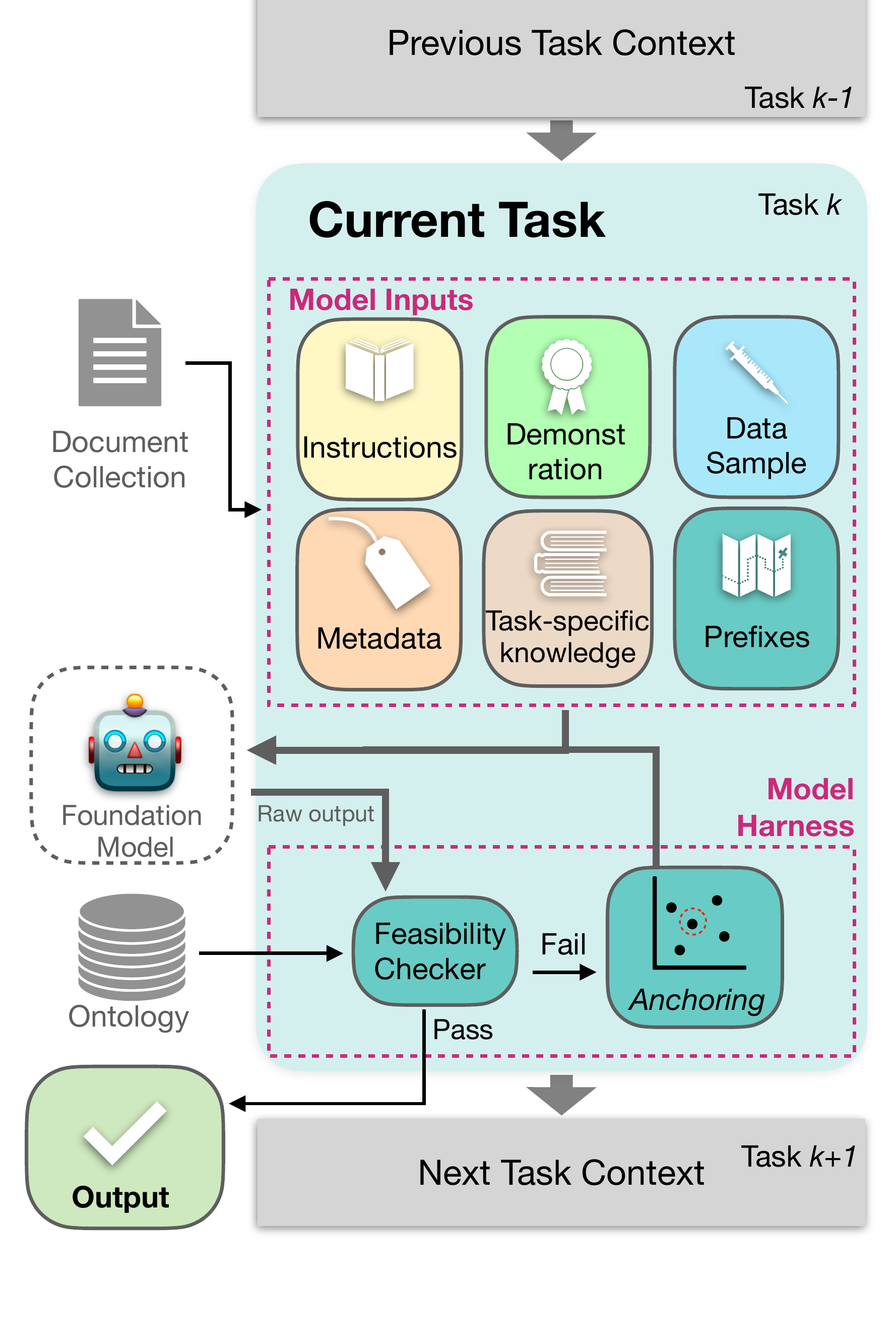}
    \caption{\textsc{Chorus} system architecture.}
    \vspace{-0.25in}
    \label{fig:architecture}
\end{figure}

Figure~\ref{fig:architecture} shows the architecture of the system. \textsc{Chorus} has a unified architecture which runs multiple tasks in the same context, allowing for information flow. Each task is run sequentially, with the output of one task fed as context into future tasks. 

For each task instance, \textsc{Chorus} generates a prompt by concatenating six inputs: context, demonstration, data samples, metadata, task-specific knowledge, and prefixes. They form the ``Model Inputs'' box in Figure~\ref{fig:architecture} and are color-coded so that they  match the colored prompt components in Figure~\ref{fig:prompts}. This natural language input is then fed to the foundation model. The output is then subject to post-processing: checks of parsability and feasibility are conducted. If these pass, the output is extracted. Otherwise, we activate a mitigation process,  called \textit{anchoring}, in order to repair the error and prevent its propagation.

\begin{figure}
{\footnotesize
\fbox{\parbox{0.75\linewidth}{\textbf{Legend}: \hlc[yellow!30]{Instruction}, \hlc[green!30]{Demonstration},  \hlc[cyan!30]{Data sample}, \hlc[orange!30]{Metadata}, \hlc[brown!30]{Task-specific knowledge}, \hlc[pink!30]{Prefixes}.}}
\vspace{0.5cm}

\begin{subfigure}[b]{\linewidth}

\texttt{\hlc[yellow!30]{For the following CSV sample, select one DBpedia.org ontology that represents the dataset from the following list:}}

\texttt{\hlc[brown!30]{AcademicJournal, AdministrativeRegion, Airline, Airport, [\ldots], University, VideoGame, Work, Wrestler.}}

\texttt{\hlc[green!30]{For example, for a dataset about hospitals, return \textasciigrave https://dbpedia.org/ontology/Hospital\textasciigrave.} \hlc[yellow!30]{Begin your answer with} \hlc[pink!30]{'https://dbpedia.org/ontology'}}.

\hlc[cyan!30]{\textasciigrave\textasciigrave\textasciigrave}\\
\hlc[orange!30]{\texttt{Brand, \textvisiblespace, ZIP, \textvisiblespace,}} \\
\hlc[cyan!30]{\texttt{Nissan, Leaf, 98112, JN1AZ0CP4C,}} \\
\hlc[cyan!30]{\texttt{Tesla, Model 3, 98074, 5YJ3E1EBXL,}} \\
\hlc[cyan!30]{[...]} \\
\hlc[cyan!30]{\textasciigrave\textasciigrave\textasciigrave}

\caption{Table-class detection}
\label{fig:prompts:t1}
\vspace{0.5cm}
\end{subfigure}
\begin{subfigure}[b]{\linewidth}
\texttt{\hlc[yellow!30]{Consider this example.} \hlc[green!30]{Input:}}

\hlc[green!30]{\textasciigrave\textasciigrave\textasciigrave}

\texttt{\hlc[green!30]{Name, Famous Book, Rk, Year}}

\texttt{\hlc[green!30]{Fyodor Dostoevsky, Crime and Punishment, 22.5, 1866}}

\texttt{\hlc[green!30]{Mark Twain, Adventures of Huckleberry Finn, 53, 1884}}

\texttt{\hlc[green!30]{Albert Camus, The Stranger, -23, 1942}}

\hlc[green!30]{\textasciigrave\textasciigrave\textasciigrave}

\texttt{\hlc[green!30]{Output:\\ \textasciigrave dbo:author, dbo:title, Unknown, dbo:releaseDate\textasciigrave.}}

\texttt{\hlc[yellow!30]{For the following CSV sample, suggest a DBPedia.org Property for each column from the \textasciigrave dbo:\textasciigrave namespace.}}

\hlc[cyan!30]{\textasciigrave\textasciigrave\textasciigrave}
\hlc[cyan!30]{\texttt{[...]}}
\hlc[cyan!30]{\textasciigrave\textasciigrave\textasciigrave}

\caption{\rone{Column-type annotation}}
\label{fig:prompts:t2}
\vspace{0.5cm}
\end{subfigure}
\begin{subfigure}[b]{\linewidth}
    \texttt{\hlc[yellow!30]{Given two Pandas Dataframes, suggest what  \textasciigrave pd.merge\textasciigrave\\ parameters to use to join the dataframes.}}

\hlc[cyan!30]{df1 =} 

\hlc[cyan!30]{\textasciigrave\textasciigrave\textasciigrave}
\hlc[cyan!30]{\texttt{[...]}}
\hlc[cyan!30]{\textasciigrave\textasciigrave\textasciigrave}

\hlc[cyan!30]{df2 =}

\hlc[cyan!30]{\textasciigrave\textasciigrave\textasciigrave}
\hlc[cyan!30]{\texttt{[...]}}
\hlc[cyan!30]{\textasciigrave\textasciigrave\textasciigrave}

\texttt{\hlc[yellow!30]{Complete the correct Pandas merge command.}} \texttt{\hlc[pink!30]{\textasciigrave pd.merge(df1, df2, left\_on=}}
\caption{Join-column prediction}
\label{fig:prompts:t3}
\end{subfigure}
}
\caption{\rone{Prompts used in this paper, materialized with examples. Most prompt elements are fixed---only the \hlc[cyan!30]{data sample} and  \hlc[orange!30]{metadata} change for each instance.}}
\label{fig:prompts}
\vspace{-0.3in}
\end{figure}

\subsection{Model Inputs}
\label{sec:approach:input}
We discuss what inputs are provided to the foundation model and how they are pre-processed and synthesized. We discuss the six components of the Model Inputs module in Figure~\ref{fig:architecture}, individually. These correspond to the six color-coded prompt components in Figure~\ref{fig:prompts}.  Once generated, all the above inputs are concatenated to into a single prompt provided to the model.

\paragraph{Instructions}  A description of the specific task (table-class detection, column-type annotation or join-column prediction) is provided to the foundation model in natural language. These are shown in yellow in Figure~\ref{fig:prompts}. For example, we translate the formal Definition~\ref{def:table:class:detection} of the first task, table-class detection, into the English sentence ``For the following CSV sample, select one DBpedia.org ontology that represents the dataset.'' For the third task, join-column prediction, we utilize a code-completion approach. We frame the task as code-completing a Pandas fragment that performs a join, with the code to complete shown in Figure~\ref{fig:prompts:t3}. We choose Pandas because it is a very popular framework, with more than millions of example lines of code on the web. This is the \textit{zero-shot prompt} setting: the model can be provided with instructions for a novel task and performs them directly.

\paragraph{Demonstration} For the first two tasks, we use the foundation models with task examples as an additional input: this is called the \textit{few-shot prompt} setting. The model is given a few demonstrations of task completion, including inputs and outputs. This is shown in Figure~\ref{fig:prompts} as green text.

\paragraph{Data sample} By serializing the input tables, we can input them into foundation models. For example, consider the example table from Figure~\ref{fig:lead_examples} in the introduction. Serializing the table allows the foundation model to ingest the data. We use the comma-separated values (\textsc{csv}) format, shown in blue in Figure~\ref{fig:prompts}.

Because the models have a limited context window size---typically in the few thousands of tokens---tables cannot always be ingested as a whole. Instead, we always serialize a sample of the rows. We find a sample size of five is sufficient. Intuitively, it suffices to consider only a few values to determine column type.

\paragraph{Metadata} Schema information including column names (headers) and keys can be incorporated into the input, above the serialized data sample. We found that foundation models can adaptively infer whether the first column of the input is a header or data row, with no modification of the input required. This is shown in orange in Figure~\ref{fig:prompts}. 

\paragraph{Task-specific knowledge} For some tasks, additional information can be used to guide the model. For \textcircled{1} table-class detection, if only certain output classes are desired, these can be listed to the model. The model will take these instructions into account when generating an output but they are not hard constraints. The encoding of such additional constraints for the table-class detection task is shown in Figure~\ref{fig:prompts:t1}.

\paragraph{Prefixes}  We also provide the model with \textit{prefixes} with which to complete. This includes the DBPedia format for the table-class detection task and a Pandas code fragment for the join-column prediction task. Both prefixes are highlighted in pink in Figure~\ref{fig:prompts}. Prefixes increase the likelihood the model will provide the output in a parsable format rather than deviating into a natural language description.

\subsection{Model Harness}
\label{sec:approach:post}
The foundation model is run within a harness that parses outputs into a symbolic representation and mitigates errors.

\paragraph{Constraint checks} Because the model is not constrained in
its outputs, it may not always output a feasible answer. In this
setting we impose three constraints: table types must belong to the
ontology classes, column types must belong to the ontology properties
and joins must be on existing columns. An output is infeasible if,
in particular, it is not parsable or if it violates any of the
three constraints. If this occurs, \textsc{chorus} performs anchoring.

\begin{figure}
    \includegraphics[width=\linewidth]{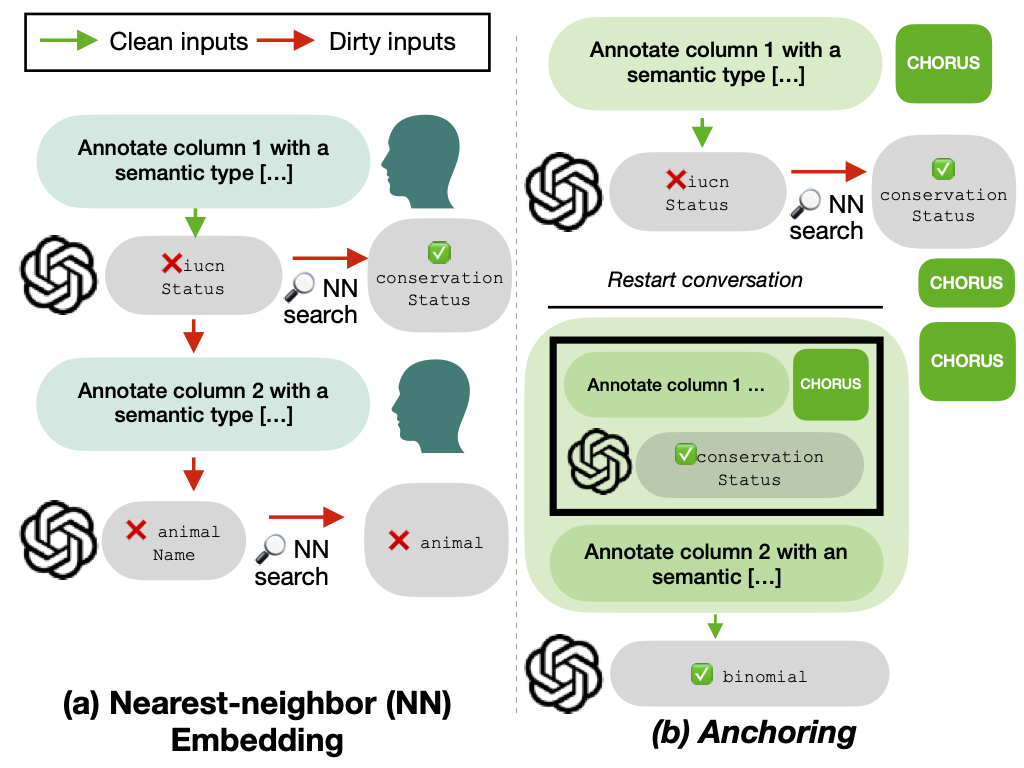}
    \caption{\textit{Anchoring} illustrated. The LLM hallucinates an imagined label, \texttt{iucnStatus}. Under the standard approach, this poisons all the upcoming tasks; the nearest-neighbor post-processing cannot recover and outputs the incorrect label \texttt{animal}. With anchoring, \textsc{chorus} intervenes when the first error is detected. A new conversation is started and a \textit{synthesized (false) history} is provided to the LLM, in which it did not make the mistake. With only clean inputs, LLM is able to correctly answer the next task correctly: \texttt{binomial}.}
    \label{fig:anchoring}
    \vspace{-0.20in}
\end{figure}

\paragraph{Anchoring} If the constraints are violated, we do not simply
move on to the next task. The risk is of hallucination snowballing~\cite{Zhang:2023aa}: once a foundation model
makes a single spurious generation, subsequent outputs are more likely to also be wrong. The model will make mistakes it would otherwise be able to avoid. For example, in Figure~\ref{fig:anchoring}(a): once nonexistent class
\texttt{iucnStatus} is suggested, another nonexistent class
\texttt{animalName} follows. Because we maintain context across tasks,
we are particularly vulnerable to this.

We call the novel domain-specific mitigation we deploy \textit{anchoring}, shown in Figure~\ref{fig:anchoring}(b). \textsc{Chorus} ends the conversation when an error is detected. It then initiates a new conversation, feeding the \textsc{llm} with a false history in which the \textsc{llm} did not hallucinate. This is possible because the conversational \textsc{llm} takes as input the full history text, which we can retroactively modify. We insert artificially an existing class from the ontology (e.g. the nearest neighbor in the embedding space to the non-existing class). Fed with this cleaner input, the model is able to directly provide the correct answer.

\section{Experiments}
\label{sec:experiments}

We empirically evaluate \textsc{chorus} on the three tasks defined in Section~\ref{sec:tasks}. For each task, we select a task-specific benchmark and compare with baselines representing the state of the art. Table-class detection \textcircled{1} is evaluated in Section~\ref{sec:experiments:t1}, \textcircled{2} column-type annotation  in Section~\ref{sec:experiments:t2}, and \textcircled{3} join-column prediction in Section~\ref{sec:experiments:t3}. The code for the experiments in available at \url{https://github.com/mkyl/CHORUS}.

\paragraph{Baselines} We considered the following state-of-the-art systems for data exploration:  relevant systems include \textsc{Tabert}~\cite{DBLP:conf/acl/YinNYR20}, \textsc{DoDuo}~\cite{Suhara:2022aa}, Sato~\cite{Zhang:2020aa}, TURL~\cite{DBLP:journals/pvldb/DengSL0020}, TaBBIE~\cite{Iida:2021aa}, Auto-suggest~\cite{DBLP:conf/sigmod/YanH20}, Trifecta Wrangler~\cite{trifacta_wrangler}, Paxata, Tableau Prep, and Sherlock. \textsc{DoDuo} outperforms TURL and Sherlock on column-type annotation~\cite{Suhara:2022aa}, so we select it for evaluation. Sato and Sherlock are similar, with Sato utilizing additional signals not found in our benchmarks, so we evaluate the better-established Sherlock. TaBBIE can embed tables but is not trained on column-type annotation unlike DoDuo and Tabert, so we avoid it for the column-type annotation task. \textsc{Tabert} is a work similar to DoDuo and TURL, but from the NLP community rather than the data management community, so we also test it too. For join-column prediction, Trifacta Wrangler outperforms Paxata and Tableau Prep~\cite{DBLP:conf/sigmod/YanH20}. Auto-Suggest is reported to outperform Trifacta Wrangler, but is a proprietary research project not released publicly. Thus we select Trifacta Wrangler for testing.

For the evaluated prior works \textsc{Tabert}, \textsc{DoDuo}, Trifecta Wrangler and Sherlock~\cite{Suhara:2022aa,trifacta_wrangler,DBLP:conf/kdd/HulsebosHBZSKDH19,DBLP:conf/acl/YinNYR20}, we utilize each tool if applicable to the task. If the baseline is not designed for a particular task, but can be straightforwardly adapted, we do so. We describe all modifications in the task subsection and always use established adaptations if available. If the modifications required would be extensive enough to become their own research project, we consider that task unsupported. In all cases, we use the pretrained embeddings without modification, as provided by the authors. Table~\ref{tab:systems-comparasion} outlines the systems we tested and tasks they support.

\begin{table}
\caption{\textit{Capabilities of related systems}. Only our system supports all studied tasks out-of-the-box and without additional training.}
\label{tab:systems-comparasion}
\begin{tabular}{lp{1.25cm}p{1.5cm}p{1.4cm}}
\toprule
System & Table-class detection & Column-type annotation & Join-column prediction \\
\midrule
\textsc{DoDuo}~\cite{Suhara:2022aa} & \centering  \textcolor{orange}{\ding{108}} & \centering \textcolor{lightgreen}{\ding{51}} & \textcolor{red}{\ding{55}} \\
\rowcolor{gray!10} \textsc{Tabert}~\cite{DBLP:conf/acl/YinNYR20} & \centering  \textcolor{orange}{\ding{108}} & \centering  \textcolor{lightgreen}{\ding{51}} & \textcolor{red}{\ding{55}} \\
Sherlock~\cite{DBLP:conf/kdd/HulsebosHBZSKDH19} & \centering \textcolor{red}{\ding{55}} & \centering \textcolor{lightgreen}{\ding{51}}  & \textcolor{red}{\ding{55}} \\
Trifacta Wrangler~\cite{trifacta_wrangler} & \centering \textcolor{red}{\ding{55}} &  \centering \textcolor{orange}{\ding{108}}  & \textcolor{lightgreen}{\ding{51}} \\
\rowcolor{gray!10} \textbf{\textsc{Chorus}} & \centering \textcolor{lightgreen}{\ding{51}} & \centering \textcolor{lightgreen}{\ding{51}} &  \textcolor{lightgreen}{\ding{51}} \\
\bottomrule
\end{tabular}\\
\textcolor{lightgreen}{\ding{51}} supported out-of-the-box,  \textcolor{red}{\ding{55}} no support \\ \textcolor{orange}{\ding{108}} required modification or training data collection (see text)
\vspace{-0.25in}
\end{table}

\textsc{DoDuo} provides two embedding variants: one trained on the WikiTables dataset and another on VizNet. We label them \textsc{DoDuo-Wiki} and \textsc{DoDuo-Viz}.

\begin{table}
\centering
\caption{\rone{Summary of the datasets used in the paper. Numbers indicate the size of the data used.}}
\begin{tabular}{lccc}
\toprule
\textbf{Dataset Title} & \# \textbf{Tables} & \textbf{Avg.} \# \textbf{Columns} & \textbf{Avg. Rows} \\
\midrule
\textsc{T2d-class} v2 & 237 & 7.41 & 118 \\
\rone{\textsc{VizNet}} & \rone{$\sim10~600$} & \rone{3.03} & \rone{5~200} \\
\rone{GitNotebooks} & \rone{24~579} & \rone{30.9} & \rone{60~242} \\
\midrule
\textbf{Overall} & \rone{35~416} & \rone{23.0} & \rone{43~491} \\
\bottomrule
\end{tabular}
\label{table:datasets}
\end{table}

\paragraph{Datasets} Table~\ref{table:datasets} outlines the three experiment benchmarks we use. For the table-class detection task, we test on the T2D-class v2 dataset~\cite{DBLP:conf/edbt/RitzeB17}, a ``gold standard'' corpus of $237$ tables, manually annotated by experts with one of $39$ DBPedia.org classes. These tables were in turn selected from the Common Crawl corpus of web tables~\cite{commoncrawl}. For column-type annotation, we sample a subset of the VizNet dataset~\cite{Hu:2019aa}, extracted by the Sherlock team~\cite{DBLP:conf/kdd/HulsebosHBZSKDH19}, comprised of \rone{$32~386$ columns with one of fifteen types from approximately $10~600$ tables}.
This is in line with prior work that uses VizNet~\cite{DBLP:journals/corr/abs-2106-07258}. For the join-column prediction task, we use a dataset we call GitNotebooks, extracted by the Auto-suggest team~\cite{DBLP:conf/sigmod/YanH20}. \rone{We select 300 tables from that dataset for which we have join data to run manually.} Here we use a sample as one of the baselines, Trifacta Wrangler, does not have an API but instead predictions must be produced manually. \rone{Separately, we run all 24 thousand tables on the baselines with an \textsc{api}.} For the first two tasks, which require defining a type system for classes and properties, we use the DBPedia ontology~\cite{DBLP:conf/lrec/MendesJB12} for our experiments. This is a community-sourced ontology and is the standard in previous studies.

\paragraph{Setup} We use the GPT-3.5 model \cite{DBLP:journals/corr/abs-2203-02155} as it is the most widely-available large model with \textsc{api} access at the time of writing. All other code was run on a commodity laptop with 8 physical \textsc{arm} cores and 16GB of main memory. \rthree{Running all experiments came to a total of \$20 in \textsc{api} costs.}

We evaluate using the metrics \textit{precision}, \textit{recall} and $F_1$ score. Precision is the proportion of true positive results out of the total predicted positive results, while recall is the proportion of true positive results out of the total actual positive results in the dataset. The $F_1$ score is the harmonic mean of precision and recall. Since we deal with a multiclass setting, we calculate these metrics for each class separately then aggregate by taking the mean, weighted by the class size. Weighted precision, recall and $F_1$ are the standard metrics in prior work~\cite{Zhang:2020aa,Suhara:2022aa,DBLP:conf/kdd/HulsebosHBZSKDH19,DBLP:journals/pvldb/CafarellaHWWZ08}. \rthree{We also report average throughput and cost for each task.}

\subsection{Table-class detection}
\label{sec:experiments:t1}
For the first task, \textcircled{1} table-class detection, we tag each table with the DBPedia ontology entry that represents the row-type of the data. \rone{Of the $1~000$ datasets that comprise the T2Dv2 dataset, $237$ tables have table-class correspondences available while 763 do not---we exclude the unlabelled ones from the supervised evaluation. We call this subset of $237$ annotated tables \textit{T2D-class v2} and use it for evaluation on this task.} We note that only $40$ classes are utilized in this ``gold standard'' mapping, while DBPedia ontology has $769$ classes.

We compare against the baselines \textsc{DoDuo} and \textsc{TaBert}. No approach in the prior work provides out-of-the-box capabilities on this task, so we add a classification layer on top of the pretrained embedding layer. After computing the column embeddings using \textsc{DoDuo} or \textsc{TaBert}, predictions are extracted by adding a pooling layer, fed to a multi-layer perceptron, and then finally taking the soft-max. This is a straightforward method of adapting the embeddings to our multi-class setting, used in prior benchmarks for table-class detection~\cite{DBLP:journals/corr/abs-2110-15132}. We fix the embeddings to their pretrained values and learn the weights of the classification layer using five-fold cross-validation.

\paragraph{Supervised variant} To allow for comparisons with prior work, we initially restrict our system to picking out of the $33$ classes. This is because all other approaches require training on labelled instances---the baselines cannot predict outside those classes. We test $33$ classes rather than $40$ because the classes that occur only once cannot be tested on baselines that require supervised training (DoDuo and TaBERT), since a result requires a disjoint training and test set.

Table~\ref{tab:table-class-results} shows the results. \textsc{Chorus} improves on the three baselines on all metrics. $F_1$ score is improved by 0.169 points, precision by $17.5$ percentage points and recall by $15.5$ percentage points. Of the baselines, DoDuo-Wiki provides the best $F_1$ and precision, while TaBERT provides the comparable recall. The best performing models, TaBERT and DoDuo-Wiki are trained on CommonCrawl, a superset of the T2Dv2 benchmark. DoDuo-Viz which is trained on the VizNet, a dataset disjoint from T2Dv2, has the weakest performance. The numbers for TaBERT are in line with prior replications~\cite{DBLP:journals/corr/abs-2110-15132}, while to the best of our knowledge this is the first benchmarking of DoDuo on this task.

\paragraph{Unsupervised variant} Next, we relax the classification domain, allowing the foundation model to choose any of the $768$ classes of the DBPedia ontology. We then compare the quality of the classes with that of the human-expert labels. DoDuo and TaBERT are not evaluated in this task setting as they cannot predict outside the classes they have observed in training.

For $93\%$ of tables, our system produces correct results. Of that portion, $83$ percentage points are comprised of exact matches, while $10$ percentage points are \textit{better-than-correct} results. This means we judge the predicted labels are clearly and unambiguously better than those selected by the benchmark authors. This is a strong claim so we list all such datasets in the technical report~\cite{Kayali:2023aa}, with evidence. For the final $6\%$ the answer is incorrect: this can mean the answer is wrong or simply worse than the label provided by the expert. This means that on the relations where \textsc{chorus} and the expert-label disagree, our system is $1.6\times$ more likely to be correct.

\rthree{\textsc{Chorus} has a throughput of nearly 31 tables per second on this benchmark and cost an average 2.5¢ per 100 table-class predictions.}

\begin{table}
    \caption{Weighted $F_1$ scores for \textit{table-class detection} on T2Dv2 dataset. Systems are compared with the expert-annotated classes for each table. The $n=237$ tables each correspond to one of $33$ \texttt{DBPedia.org} classes.}
    \centering
\begin{tabular}{lccc}
\toprule

& \textbf{$F_1$-score} & \textbf{Precision} & \textbf{Recall} \\
\midrule
DoDuo-Viz & 0.654 & 66.8\% & 68.3\% \\
DoDuo-Wiki & 0.757 & 78.6\% & 76.9\% \\
TaBERT& 0.746 & 76.3\% & 76.8\% \\
\textbf{\textsc{Chorus}} & \textbf{0.926} & \textbf{96.1\%} & \textbf{92.4}\% \\
\bottomrule
\end{tabular}
    \label{tab:table-class-results}
\end{table}

\subsection{Column-type annotation}
\label{sec:experiments:t2}
Next, we compare the ability of our system to assign classes to table columns. \textsc{VizNet} is a collection of tables, extracted by the Sherlock~\cite{DBLP:conf/kdd/HulsebosHBZSKDH19} team from the VizNet repository~\cite{Hu:2019aa} of data visualizations and open datasets. \rone{VizNet comprises 31 million columns in total, of which a test set of 142~000 can be used for evaluation---the rest have trained on by DoDuo and Sherlock. 
Of those, we select a subset of $15$ classes which are supported by both DoDuo-Wiki and DoDuo-VizNet (these baselines support a disjoint set of classes), arriving at $32~386$ test columns used as a benchmark in this section.}

\paragraph{Baselines} We compare against TaBERT~\cite{DBLP:conf/acl/YinNYR20}, DoDuo~\cite{Suhara:2022aa} and Sherlock~\cite{DBLP:conf/kdd/HulsebosHBZSKDH19} on this task. Since Sherlock and DoDuo are designed for column annotation, we use the out-of-the-box model provided by the original teams. \rone{We restrict both to the fifteen target classes by setting the probabilities of non-target classes to zero. For DoDuo-Wiki, which supports a distinct set of classes, we perform a manual mapping to the class names used by DoDuo-VizNet and Sherlock.} For TaBERT we train an additional classification layer on top of the pre-trained embeddings that these frameworks provide. We fix the embeddings to their pretrained values and learn the weights of the classification layer using five-fold cross-validation.

\paragraph{Results} Table~\ref{tab:viznet-class-results} contains the results for the \textsc{VizNet} dataset. Our \textsc{fm}-based approach improves performance on the measured metrics of $F_1$-score, precision and recall. The best performing method is Sherlock, narrowly beating DoDuo-VizNet, with a \rone{0.954 $F_1$ score}. If we consider methods which are not specifically pretrained on VizNet (note, which is also the test set) \textsc{chorus} is the best performing on all three metrics. It has comparable $F_1$ and precision to Sherlock, but $6$ percentage points lower recall.

Note in particular DoDuo-Wiki, which does not have access to VizNet at pretraining time, has a large regression in performance compared to DoDuo-Viznet, nearly half $F_1$ points. This drop is in line with previous results, see Section~\ref{sec:discussion}. We sanity-check the low scores of TaBERT by replicating previously reported scores~\cite{DBLP:journals/corr/abs-2110-15132}.

\rthree{\textsc{Chorus} achieves a competitive throughput of 41 columns per second (col/s), comparable to Sherlock's 50 col/s and exceeding DoDuo's 7.3 col/s and TaBERT's 4.5 col/s. This corresponds to benchmark completion in 13 minutes, as contrasted with over 2 hours 10 minutes for TaBERT. The average cost of GPT-3.5 calls for this task was 1.3¢ cents per 100 columns.}

\begin{table}
    \caption{\rone{Weighted $F_1$ scores for \textit{column-type annotation} on \textsc{VizNet} test set, with $n=32~000$ columns. Systems are compared with the ``gold standard'' classes for each column. Methods which are also pre-trained on \textsc{VizNet} are marked with an asterisk $\ast$.}}
    \centering
\begin{tabular}{lccc}
\toprule
  & \textbf{$F_1$-score} & \textbf{Precision} & \textbf{Recall} \\ \midrule
DoDuo-VizNet$\ast$  & \rone{0.876} & \rone{89.4\%} & \rone{87.2\%} \\
Sherlock$\ast$  & \rone{0.954} & \rone{96.2\%} & \rone{94.6\%} \\ \midrule
TaBERT & \rone{0.321} & \rone{32.6\%} & \rone{32.0\%} \\
DoDuo-Wiki  & \rone{0.440} & \rone{59.2\%} & \rone{45.4\%} \\
\textbf{\textsc{Chorus}} & \textbf{\rone{0.891}} & \textbf{\rone{91.2\%}} & \textbf{\rone{88.8\%}} \\
\bottomrule
\end{tabular}
    \label{tab:viznet-class-results}
    \vspace{-0.1in}
\end{table}

\subsection{Join-column prediction}
\label{sec:experiments:t3}
Finally, we evaluate our approach's ability to suggest which columns are the correct choice for a join, the join-column prediction task. We use the \textit{GitNotebooks} dataset from \cite{DBLP:conf/sigmod/YanH20}, a collection of 4 million Python notebooks (and their associated relational tables) including 24 thousand joins collected from Github. One of the baselines, Trifacta Wrangler, requires manual execution and recording of each prediction. For that reason  we restrict this benchmark to 300 randomly sampled tables.

\paragraph{Baselines} For this task, we compare with three baselines. Jaccard similarity, $J$, is the first. Two columns are selected such that $\operatorname*{argmax}_{c \in C^T, c' \in C^{T'}} J(c, c')$ where $J(X, Y) = |X \cap Y| / |X \cup Y|$. This is a commonly used approach in the literature~\cite{DBLP:journals/pvldb/ChenNC14, DBLP:conf/sigmod/DasuJMS02, DBLP:journals/tkde/NargesianPBZM23,DBLP:conf/sigmod/YanH20}. Another baseline is Levenshtein distance~\cite{levenshtein1966binary}, which selects the pair of column names with the smallest edit distance between them. The final baseline is Trifacta Wrangler~\cite{trifacta_wrangler}, a commercial product spun off from the Wrangler research line~\cite{DBLP:conf/chi/KandelPHH11}. When joining two tables in this product, it suggests the keys on which to join them. As no \textsc{api} was available, we obtain all Trifacta predictions by joining manually.

\paragraph{Results} Table~\ref{tab:join-column-results} shows the quality of estimates for our approach and the baselines. We measure the quality of the predictions by the same criteria as the previous tasks. By these metrics, our approach improves the quality of predictions and beats the next-best approach by a clear margin: $F_1$ score is improved by $0.072$, precision by $8.4$ percentage points and recall by $6.0$ percentage points. \rone{This performance is maintained when scaling to the full dataset.} \rthree{On this task, our system has an average throughput of 23.5 predictions per second and cost approximately 5¢ cents per 100 predicted joins.}

\begin{table}
    \caption{\rone{$F_1$ scores, precision and recall for the \textit{join-column prediction} task on GitNotebooks dataset.}}
    \centering
\begin{tabular}{lccc}
\toprule
 & \textbf{$F_1$-score} & \textbf{Precision} & \textbf{Recall} \\
\midrule
Manually-run subset, $n=300$ & & & \\
\quad Jaccard & 0.575 & 60.7\% & 54.7\% \\
\quad Levenshtein & 0.718 & 72.3\% & 71.3\% \\
\quad Trifacta Wrangler & 0.823 & 82.6\% & 82.0\% \\
\quad \textbf{\textsc{Chorus}}  &  \textbf{0.895} & \textbf{91.0}\% & \textbf{88.0}\% \\
\midrule
\rone{Full dataset, $n=24~579$} & & & \\
\quad \rone{Jaccard} & \rone{0.458} & \rone{63.3\%} & \rone{35.9\%}  \\
\quad \rone{Levenshtein} & \rone{0.777} & \rone{78.7\%} &  \rone{76.8\%} \\
\quad \rone{Trifacta Wrangler} & \multicolumn{3}{c}{\rone{No \textsc{api}}} \\
\quad \rone{\textbf{\textsc{Chorus}}} & \rone{\textbf{0.912}} & \rone{\textbf{93.2\%}} & \rone{\textbf{89.4\%}} \\
\bottomrule
\end{tabular}
    \label{tab:join-column-results}
\end{table}

\subsection{Dataset contamination}
\label{sec:experiments:contamination}
Here we perform an experiment to validate whether any of the testing data occurred in the training corpus of the large-language model, an issue called \textit{dataset contamination} or \textit{data leakage}. Because these models are trained on internet data~\cite{DBLP:journals/corr/abs-2101-00027} and we use public benchmarks, they may have seen the test data in training.

We test on seven guaranteed-unseen tables (listed in the technical report~\cite{Kayali:2023aa}) and their columns, all uploaded between April--June 2023 to the federal data repository Data.gov. They are guaranteed-unseen because the foundation model training was completed on or before March 2023.  Repeating the supervised column-type annotation task as in Section~\ref{sec:experiments:t2}, we measure a 0.857 $F_1$ score, 90.0\% precision and 81.8\% recall. This is within $0.01$ $F_1$ points, 0.1\% precision and 5\% recall of the benchmark results. See Table~\ref{tab:contamination}. The recall falling more is expected, as this reflects
the datasets being more diverse and therefore difficult to classify.

\begin{table}
    \caption{Data contamination experiment. Weighted $F_1$ scores for \textit{table-class detection} on public benchmarks versus tables the foundation model is guaranteed to have not been trained on. }
    \centering
\begin{tabular}{lccc}
\toprule
\textbf{Dataset} & \textbf{$F_1$-score} & \textbf{Precision} & \textbf{Recall} \\
\midrule
Public benchmark (VizNet) & 0.865 & 90.1\% & 86.7\% \\
Guaranteed-unseen & 0.857 & 90.0\% & 81.8\% \\
\bottomrule
\end{tabular}
    \label{tab:contamination}
    
\end{table}

\subsection{System characteristics}
\label{sec:experiments:misc}
\begin{figure}
    \centering
    \includegraphics[width=0.85\linewidth]{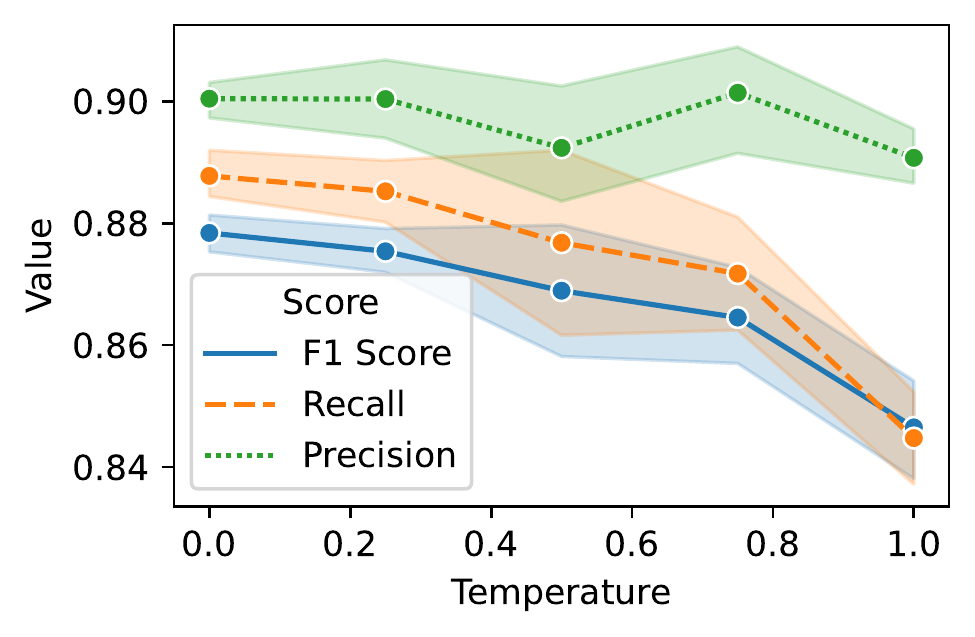}
    \caption{\textit{Determinism vs. performance}. We conduct 25 runs of \textsc{chorus} on the T2D table class benchmark. Shaded bands indicate confidence intervals. Temperature is a parameter controlling the randomness of the foundation model, with zero being the most (but not completely) deterministic.}
    \label{fig:temperature}
    \vspace{-0.2in}
\end{figure}

\paragraph{Determinism} We examine the impact of nondeterminism in the foundation model on the performance of \textsc{chorus}. The randomness of the generation is controlled by the \textit{temperature} hyperparameter. To assure that the results of \textsc{chorus} are reliable, we conduct the following experiment: we run the T2D table-class detection benchmark 25 times, five trials for each value of $T$ between $0, 1/4, ..., 1$.
Figure~\ref{fig:temperature} shows the result. \textsc{Chorus}'s performance is consistent: at the ideal temperature setting the $F_1$ score sees error bars of 0.01 $F_1$ points. The best performance is obtained at lowest temperature, $0.0$---this is in contrast to NLP tasks like summarization that benefit from higher temperatures. In the prior experiments we use the default temperature, as to get \textsc{chorus} running with minimal hyperparameter tuning.

\paragraph{Alternative models} To demonstrate the versatility of this approach, we run \textsc{chorus} with three alternative, open-source foundation models on the table-class detection task. We consider Vicuna~\cite{Zheng:2023aa}, a variant of LLaMA~\cite{Touvron:2023aa} at two sizes: 13 billion parameters and 7 billion parameters. The more advanced model is LLaMA 2~\cite{Touvron:2023ab}, the SOTA open-source model with 70 billion parameters.

Table~\ref{tab:llama} shows the results. While OpenAI's GPT model performs best, the best open-source model is very competitive. LLaMA 2 outperforms the best baseline model for this task---DoDuo-Wiki---by 0.136 $F_1$ points, on precision by 13.6 percentage points and on recall by 9.6 percentage points. This model lags behind the proprietary and larger GPT model by only a modest 0.03 $F_1$ points. Open-source LLMs are now compelling alternatives on the tested task.

\begin{table}
    \caption{Alternative foundation models. Weighted $F_1$ scores for \textit{table-class detection} on T2Dv2 dataset, for different choices of  foundation model used by \textsc{chorus}. Parameter size in brackets. GPT-3.5 numbers identical to experiment in Figure~\ref{tab:table-class-results}.}
    \centering
\begin{tabular}{lccc}
\toprule
 & \multicolumn{3}{c}{\textbf{Table-class correctness}} \\
\cmidrule(lr){2-4}
\textbf{Model choice} & \textbf{$F_1$-score} & \textbf{Precision} & \textbf{Recall} \\
\midrule
GPT-3.5 (175B) & 0.926 & 96.1\% & 92.4\% \\
LLaMA 2 (70B) & 0.893 & 92.2\% & 86.5\% \\
Vicuna/LLaMA (13B) & 0.713 & 79.2\% & 64.1\% \\
Vicuna/LLaMA (7B) & 0.713 & 75.3\% & 67.5\% \\
\bottomrule
\end{tabular}
    \label{tab:llama}
\end{table}

\begin{figure}
    \centering
    \includegraphics[width=0.85\linewidth]{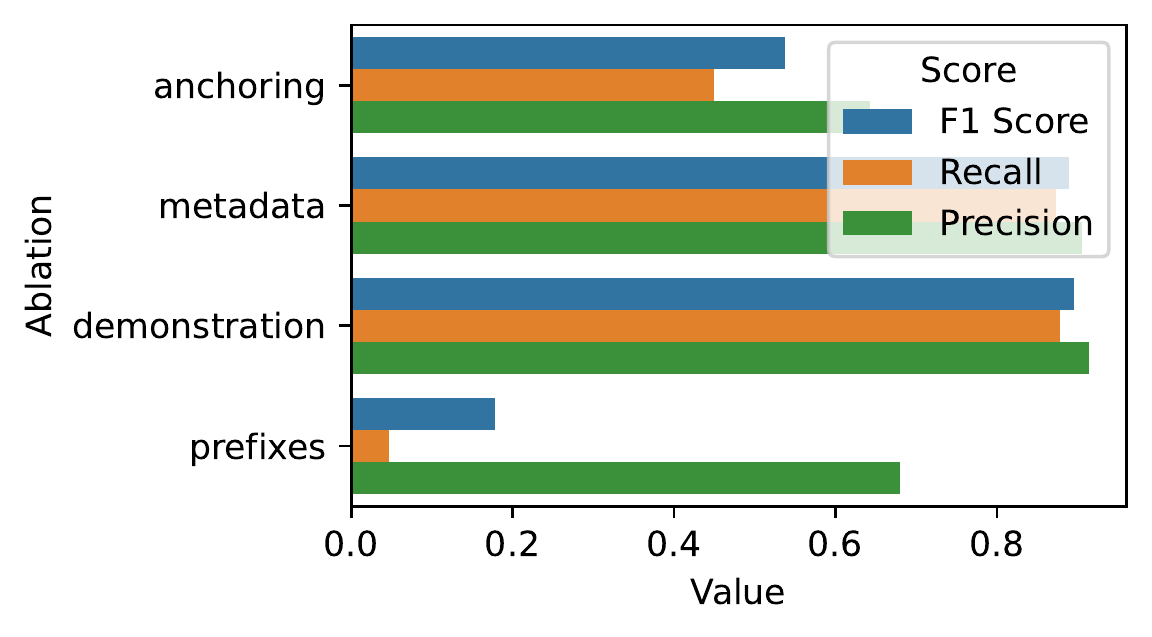}
    \caption{\textit{Ablation experiments}. We ablate key features of \textsc{Chorus} and report performance characteristics on the T2D table-class detection task.}
    \label{fig:ablations}
\end{figure}

\paragraph{Ablations} We conduct ablation experiments to measure the contribution of individual components of \textsc{chorus}. We remove one component at a time and note the loss of scores compared to the unaltered model. Figure~\ref{fig:ablations} shows the results. First, we remove the demonstration from the prompt. This results in an F1 score loss of 0.03, a recall loss of 4.7 and a precision loss of 4.7 percentage points. Next, we remove the metadata where it is available. This results in a cumulative F1 score loss of 0.04, a recall loss of 5.1 and a precision loss of 5.6 percentage points. \rtwo{After that, we disable anchoring. This results in a cumulative F1 score loss of 0.389, a recall loss of 47.4 and a precision loss of 31.9 percentage points. From this the prevalence of hallucinations can be gleaned: the substantive score loss implies that hallucination is highly prevalent. The score improvement with anchoring implies that it can be recovered from: the \textsc{fm} mistakes are not severe but rather resemble the ground truth.} Finally, we remove the prefixes from the prompt. This results in a cumulative F1 score loss of 0.736, a recall loss of 92.3 and a precision loss of 53.1 percentage points.

\section{Discussion}
\label{sec:discussion}

\paragraph{Training data collection} A major advantage of a foundation-model approach is that there is no need for training on specific tasks. In contrast, \textsc{TaBERT} requires 26 million tables for training its embeddings. In \cite{DBLP:journals/pvldb/DengSL0020}, the use of 250 labels for one task is considered a ``small dataset'' by the authors and leads to subpar performance. In contrast, our prompts in Figure~\ref{fig:prompts} use zero or none examples for each task.

\paragraph{Out-of-domain performance} We note a troubling pattern of a lack of cross-domain generalization in representation-learning approaches. The tested baselines degrade when used to embed tables not from the dataset the embeddings were trained on.
This finding is in-line with prior work: regressions of up to 0.40 and 0.30 $F_1$ points when generalizing to new datasets have been reported \cite{DBLP:journals/pvldb/DengSL0020, DBLP:journals/corr/abs-2106-07258}.

\paragraph{Flexibility} Another advantage of \textsc{chorus} we observe in the experiments is task adaptability. In the \textcircled{1} table-class detection task, we are able to switch the prediction domain easily. Restricting to the 33 classes used by the benchmark can be done by providing the permitted classes to the foundation model; allowing the model to generalize to other DBPedia classes (the \textit{unsupervised} heading of Section~\ref{sec:experiments:t1}) is as simple as omitting those instructions. Contextual information, such as table title or \textsc{url}, could be as easily added. Previously, such modifications would require retraining the embeddings.

\paragraph{Limitations and risks} We control the risk of dataset contamination by testing for it in Section~\ref{sec:experiments:contamination}. The performance of \textsc{chorus} on guaranteed-unseen datasets is comparable to those in public benchmarks, so good performance on the those benchmarks cannot be explained away as simple data contamination. Separately, formal linguistic fluency means that errors may fool human reviewers~\cite{langer1978mindlessness, Langford:1999aa}. This has been called \textit{subtle misinformation} in prior work~\cite{Rae:2021aa}. Finally, prompts may not be robust to changes~\cite{Zhao:2021aa}.
\rthree{\paragraph{Pattern-matching hypothesis} We posit that \textsc{llm}s outperform representation learning approaches because they are able to extract a large number of linear transformations over the embedding space from their training data and compose them in simple ways. In the full technical report, we conduct an experiment contrasting \textsc{llm} activations with \textsc{doduo} weights, supporting this hypothesis~\cite{Kayali:2023aa}. Prior work~\cite{Olsson:2022aa} supports this hypothesis and the limits of \textsc{llms} in composing these patterns, generally failing to combine patterns after about four reasoning steps~\cite{Dziri:2023aa}.}
\\
\balance

\textbf{Future directions:} \rtwo{\textit{Additional tasks.} The above hypothesis suggests the promising performance may extend to many more tasks. Related tasks to be explored include \textit{schema auto-completion}~\cite{DBLP:journals/pvldb/CafarellaHWWZ08}, where missing parts of a partial schema are suggested to the user; \textit{join-graph traversal}, where successive tables to join on are suggested~\cite{Dong:2022aa}; and outlier detection, where erroneous data are detected. These are all promising because they involve composing simple patterns that are prevalent in \textsc{fm}'s training data. On the other hand, novel approaches will likely be needed to apply \textsc{fm}s to tasks like \textit{data provenance}~\cite{10.1145/1265530.1265535}, since these involving tracking more patterns than \textsc{fm}'s are thought to currently support. Incorporating proprietary information may also be difficult to due the dearth of these patterns in the training data.}

\paragraph{Private or domain-specific datasets} As with all the tested baselines, the foundation models are trained on public data. The distribution of data in the public sphere differs significantly from that in specialized domains or private data. It is worth investigating whether the  observed capabilities continue to  hold on e.g. enterprise data lakes. Further application to domain-specific ontology such as \textsc{dron}, a pharmaceutical ontology of drugs, would also be interesting to investigate.

\paragraph{Fine-tuned or distilled models} Larger models are of interest because if scaling laws continue to hold, their performance should improve~\cite{Wei:2022ab}. The large inference cost of foundation models has sparked interest in model distillation: reducing the size of the model without incurring a reduction in specific capabilities~\cite{West:2022aa, Dettmers:2023aa} but at cost of degraded emergent abilities~\cite{Gudibande:2023aa}.
\section{Related Work}
\label{sec:related}

The seminal early work is WebTables~\cite{DBLP:journals/pvldb/CafarellaHWWZ08}, which extracts relational tables from web data, annotated with metadata for discoverability. This work introduced the related tasks: \textit{schema auto-completion}, \textit{attribute synonym finding}, and \textit{join-graph traversal}. Early work on \textit{wrapper induction}~\cite{DBLP:conf/ijcai/KushmerickWD97} also extracted tables from heterogeneous sources.

The promise of foundation models for data profiling was outlined in a recent position paper~\cite{Trummer:2022aa}. This paper was based on evidence of foundation models being able to predict correlations in data from the column names~\cite{Trummer:2021aa}. Another work considered foundation models for data wrangling~\cite{Narayan:2022ab}: comprising the tasks of entity matching, error detection and data imputation. Finally, most recently foundation models have been applied to the classic problem of wrapper induction in the system \textsc{evaporate}~\cite{Arora:2023aa}.

The currently deployed generation of approaches has focused on representation learning. These include \textsc{turl}~\cite{DBLP:journals/pvldb/DengSL0020}, \textsc{t}a\textsc{bert}~\cite{DBLP:conf/acl/YinNYR20}, DoDuo~\cite{Suhara:2022aa} and \textsc{tabbie}~\cite{Iida:2021aa}. These explore the use of fine-tuned language models for similar tasks. Prior to these table-embedding approaches, the prior generation of data tools involved data-intensive deep learning for specific tasks, e.g. Sherlock~\cite{DBLP:conf/kdd/HulsebosHBZSKDH19} and  Sato~\cite{Zhang:2020aa}.

Data discovery within data lakes is an active area of research, with recent works including: unionability search~\cite{Khatiwada:2023aa}, joinability search~\cite{Zhu:2019aa}, new index structures for faster correlated dataset search~\cite{Santos:2022aa} and end-to-end systems for data ingestion and profiling~\cite{Castelo:2021aa}. Recent tutorials~\cite{DBLP:journals/pvldb/NargesianZMPA19, Fan:2023aa} outline the prevalence of the problem of unstructured document data management. A user-study of scientists conclude that ``current systems fail to sufficiently support scientists in their data-seeking process''~\cite{papenmeier2021genuine}. One dataset-search survey~\cite{DBLP:journals/vldb/ChapmanSKKIKG20} highlights key open problems: more natural query languages, better data integration, and incorporating external knowledge.

Commercial products derived from this area of research include and include Microsoft Azure Data Lake, Databricks Lakehouse~\cite{DBLP:journals/pvldb/CafarellaHLMYWW18}, Trifacta Wrangler~\cite{trifacta_wrangler}, Tableau Prep and Paxata. Industry-led prototypes include Sigma Computing's WarpGate~\cite{DBLP:journals/corr/abs-2212-14155} and Google Research's Dataset Search~\cite{DBLP:conf/www/BrickleyBN19}.

\section{Conclusion}
\label{sec:conclusions}
We propose \textsc{chorus} to integrate foundation models for data discovery. We show it provides superior performance on three exemplars: table-class annotation, column-type detection and join-column prediction. Experimentally, we find \textsc{Chorus} more robust than prior representation-learning approaches on a variety of benchmark datasets. We conclude that foundation models hold promise as a core component of next generation data discovery systems.
\begin{acks}
Our thanks to Yejin Choi for guidance on and productive discussions about language models. We are grateful to the authors of TaBERT, Sherlock and DoDuo for high-quality, replicable experiment code and documentation. We appreciate Cynthia Richey's and Kyle Deeds' assistance with the manuscript. Finally, thanks to Magdalena Balazinska for counsel during manuscript writing. This material is based upon work supported by the National Science Foundation under Grants NSF-BSF 2109922 and NSF IIS 2314527.
\end{acks}
\bibliography{bibliography}

\begin{thebibliography}{77}
\providecommand{\natexlab}[1]{#1}
\providecommand{\url}[1]{\texttt{#1}}
\expandafter\ifx\csname urlstyle\endcsname\relax
  \providecommand{\doi}[1]{doi: #1}\else
  \providecommand{\doi}{doi: \begingroup \urlstyle{rm}\Url}\fi

\bibitem[Abdelmageed et~al.(2022)Abdelmageed, Chen, Cutrona, Efthymiou,
  Hassanzadeh, Hulsebos, Jim{\'{e}}nez{-}Ruiz, Sequeda, and
  Srinivas]{DBLP:conf/semweb/AbdelmageedCCEH22}
Nora Abdelmageed, Jiaoyan Chen, Vincenzo Cutrona, Vasilis Efthymiou, Oktie
  Hassanzadeh, Madelon Hulsebos, Ernesto Jim{\'{e}}nez{-}Ruiz, Juan Sequeda,
  and Kavitha Srinivas.
\newblock Results of semtab 2022.
\newblock In Vasilis Efthymiou, Ernesto Jim{\'{e}}nez{-}Ruiz, Jiaoyan Chen,
  Vincenzo Cutrona, Oktie Hassanzadeh, Juan Sequeda, Kavitha Srinivas, Nora
  Abdelmageed, and Madelon Hulsebos, editors, \emph{Proceedings of the Semantic
  Web Challenge on Tabular Data to Knowledge Graph Matching, SemTab 2021,
  co-located with the 21st International Semantic Web Conference, {ISWC} 2022,
  Virtual conference, October 23-27, 2022}, volume 3320 of \emph{{CEUR}
  Workshop Proceedings}, pages 1--13. CEUR-WS.org, 2022.
\newblock URL \url{https://ceur-ws.org/Vol-3320/paper0.pdf}.

\bibitem[Anaconda(2021)]{Anaconda:2021aa}
Inc. Anaconda.
\newblock State of data science.
\newblock
  https://www.anaconda.com/resources/whitepapers/state-of-data-science-2021,
  July 2021.

\bibitem[Andreas(2022)]{Andreas:2022aa}
Jacob Andreas.
\newblock Language models as agent models, 2022.

\bibitem[Arora et~al.(2023)Arora, Yang, Eyuboglu, Narayan, Hojel, Trummer, and
  R{\'e}]{Arora:2023aa}
Simran Arora, Brandon Yang, Sabri Eyuboglu, Avanika Narayan, Andrew Hojel,
  Immanuel Trummer, and Christopher R{\'e}.
\newblock Language models enable simple systems for generating structured views
  of heterogeneous data lakes, 2023.

\bibitem[Bommasani et~al.(2021)Bommasani, Hudson, Adeli, Altman, Arora, von
  Arx, Bernstein, Bohg, Bosselut, Brunskill, Brynjolfsson, Buch, Card,
  Castellon, Chatterji, Chen, Creel, Davis, Demszky, Donahue, Doumbouya,
  Durmus, Ermon, Etchemendy, Ethayarajh, Fei{-}Fei, Finn, Gale, Gillespie,
  Goel, Goodman, Grossman, Guha, Hashimoto, Henderson, Hewitt, Ho, Hong, Hsu,
  Huang, Icard, Jain, Jurafsky, Kalluri, Karamcheti, Keeling, Khani, Khattab,
  Koh, Krass, Krishna, Kuditipudi, and
  et~al.]{DBLP:journals/corr/abs-2108-07258}
Rishi Bommasani, Drew~A. Hudson, Ehsan Adeli, Russ~B. Altman, Simran Arora,
  Sydney von Arx, Michael~S. Bernstein, Jeannette Bohg, Antoine Bosselut, Emma
  Brunskill, Erik Brynjolfsson, Shyamal Buch, Dallas Card, Rodrigo Castellon,
  Niladri~S. Chatterji, Annie~S. Chen, Kathleen Creel, Jared~Quincy Davis,
  Dorottya Demszky, Chris Donahue, Moussa Doumbouya, Esin Durmus, Stefano
  Ermon, John Etchemendy, Kawin Ethayarajh, Li~Fei{-}Fei, Chelsea Finn, Trevor
  Gale, Lauren Gillespie, Karan Goel, Noah~D. Goodman, Shelby Grossman, Neel
  Guha, Tatsunori Hashimoto, Peter Henderson, John Hewitt, Daniel~E. Ho, Jenny
  Hong, Kyle Hsu, Jing Huang, Thomas Icard, Saahil Jain, Dan Jurafsky,
  Pratyusha Kalluri, Siddharth Karamcheti, Geoff Keeling, Fereshte Khani, Omar
  Khattab, Pang~Wei Koh, Mark~S. Krass, Ranjay Krishna, Rohith Kuditipudi, and
  et~al.
\newblock On the opportunities and risks of foundation models.
\newblock \emph{CoRR}, abs/2108.07258, 2021.
\newblock URL \url{https://arxiv.org/abs/2108.07258}.

\bibitem[Brickley et~al.(2019)Brickley, Burgess, and
  Noy]{DBLP:conf/www/BrickleyBN19}
Dan Brickley, Matthew Burgess, and Natasha~F. Noy.
\newblock Google dataset search: Building a search engine for datasets in an
  open web ecosystem.
\newblock In Ling Liu, Ryen~W. White, Amin Mantrach, Fabrizio Silvestri,
  Julian~J. McAuley, Ricardo Baeza{-}Yates, and Leila Zia, editors, \emph{The
  World Wide Web Conference, {WWW} 2019, San Francisco, CA, USA, May 13-17,
  2019}, pages 1365--1375. {ACM}, 2019.
\newblock \doi{10.1145/3308558.3313685}.
\newblock URL \url{https://doi.org/10.1145/3308558.3313685}.

\bibitem[Bubeck et~al.(2023)Bubeck, Chandrasekaran, Eldan, Gehrke, Horvitz,
  Kamar, Lee, Lee, Li, Lundberg, Nori, Palangi, Ribeiro, and
  Zhang]{bubeck2023sparks}
S{\'e}bastien Bubeck, Varun Chandrasekaran, Ronen Eldan, Johannes Gehrke, Eric
  Horvitz, Ece Kamar, Peter Lee, Yin~Tat Lee, Yuanzhi Li, Scott Lundberg,
  Harsha Nori, Hamid Palangi, Marco~Tulio Ribeiro, and Yi~Zhang.
\newblock Sparks of artificial general intelligence: Early experiments with
  gpt-4, 2023.

\bibitem[Cafarella et~al.(2008)Cafarella, Halevy, Wang, Wu, and
  Zhang]{DBLP:journals/pvldb/CafarellaHWWZ08}
Michael~J. Cafarella, Alon~Y. Halevy, Daisy~Zhe Wang, Eugene Wu, and Yang
  Zhang.
\newblock Webtables: exploring the power of tables on the web.
\newblock \emph{Proc. {VLDB} Endow.}, 1\penalty0 (1):\penalty0 538--549, 2008.
\newblock \doi{10.14778/1453856.1453916}.
\newblock URL \url{http://www.vldb.org/pvldb/vol1/1453916.pdf}.

\bibitem[Cafarella et~al.(2018)Cafarella, Halevy, Lee, Madhavan, Yu, Wang, and
  Wu]{DBLP:journals/pvldb/CafarellaHLMYWW18}
Michael~J. Cafarella, Alon~Y. Halevy, Hongrae Lee, Jayant Madhavan, Cong Yu,
  Daisy~Zhe Wang, and Eugene Wu.
\newblock Ten years of webtables.
\newblock \emph{Proc. {VLDB} Endow.}, 11\penalty0 (12):\penalty0 2140--2149,
  2018.
\newblock \doi{10.14778/3229863.3240492}.
\newblock URL \url{http://www.vldb.org/pvldb/vol11/p2140-cafarella.pdf}.

\bibitem[Castelo et~al.(2021)Castelo, Rampin, Santos, Bessa, Chirigati, and
  Freire]{Castelo:2021aa}
Sonia Castelo, R{\'{e}}mi Rampin, A{\'{e}}cio S.~R. Santos, Aline Bessa,
  Fernando Chirigati, and Juliana Freire.
\newblock Auctus: {A} dataset search engine for data discovery and
  augmentation.
\newblock \emph{Proc. {VLDB} Endow.}, 14\penalty0 (12):\penalty0 2791--2794,
  2021.
\newblock \doi{10.14778/3476311.3476346}.
\newblock URL \url{http://www.vldb.org/pvldb/vol14/p2791-castelo.pdf}.

\bibitem[Chapman et~al.(2020)Chapman, Simperl, Koesten, Konstantinidis,
  Ib{\'{a}}{\~{n}}ez, Kacprzak, and Groth]{DBLP:journals/vldb/ChapmanSKKIKG20}
Adriane Chapman, Elena Simperl, Laura Koesten, George Konstantinidis,
  Luis{-}Daniel Ib{\'{a}}{\~{n}}ez, Emilia Kacprzak, and Paul Groth.
\newblock Dataset search: a survey.
\newblock \emph{{VLDB} J.}, 29\penalty0 (1):\penalty0 251--272, 2020.
\newblock \doi{10.1007/s00778-019-00564-x}.
\newblock URL \url{https://doi.org/10.1007/s00778-019-00564-x}.

\bibitem[Chen et~al.(2014)Chen, Narasayya, and
  Chaudhuri]{DBLP:journals/pvldb/ChenNC14}
Zhimin Chen, Vivek~R. Narasayya, and Surajit Chaudhuri.
\newblock Fast foreign-key detection in microsoft {SQL} server powerpivot for
  excel.
\newblock \emph{Proc. {VLDB} Endow.}, 7\penalty0 (13):\penalty0 1417--1428,
  2014.
\newblock \doi{10.14778/2733004.2733014}.
\newblock URL \url{http://www.vldb.org/pvldb/vol7/p1417-chen.pdf}.

\bibitem[Cong et~al.(2022)Cong, Gale, Frantz, Jagadish, and
  Demiralp]{DBLP:journals/corr/abs-2212-14155}
Tianji Cong, James Gale, Jason Frantz, H.~V. Jagadish, and {\c{C}}agatay
  Demiralp.
\newblock Warpgate: {A} semantic join discovery system for cloud data
  warehouses.
\newblock \emph{CoRR}, abs/2212.14155, 2022.
\newblock \doi{10.48550/arXiv.2212.14155}.
\newblock URL \url{https://doi.org/10.48550/arXiv.2212.14155}.

\bibitem[Court(2023)]{Court:2023aa}
District Court.
\newblock Mata v. avianca, inc. (1:22-cv-01461).
\newblock Southern District of New York, New York, June 2023.

\bibitem[Dasu et~al.(2002)Dasu, Johnson, Muthukrishnan, and
  Shkapenyuk]{DBLP:conf/sigmod/DasuJMS02}
Tamraparni Dasu, Theodore Johnson, S.~Muthukrishnan, and Vladislav Shkapenyuk.
\newblock Mining database structure; or, how to build a data quality browser.
\newblock In Michael~J. Franklin, Bongki Moon, and Anastassia Ailamaki,
  editors, \emph{Proceedings of the 2002 {ACM} {SIGMOD} International
  Conference on Management of Data, Madison, Wisconsin, USA, June 3-6, 2002},
  pages 240--251. {ACM}, 2002.
\newblock \doi{10.1145/564691.564719}.
\newblock URL \url{https://doi.org/10.1145/564691.564719}.

\bibitem[Deng et~al.(2020)Deng, Sun, Lees, Wu, and
  Yu]{DBLP:journals/pvldb/DengSL0020}
Xiang Deng, Huan Sun, Alyssa Lees, You Wu, and Cong Yu.
\newblock {TURL:} table understanding through representation learning.
\newblock \emph{Proc. {VLDB} Endow.}, 14\penalty0 (3):\penalty0 307--319, 2020.
\newblock \doi{10.5555/3430915.3442430}.
\newblock URL \url{http://www.vldb.org/pvldb/vol14/p307-deng.pdf}.

\bibitem[Dettmers et~al.(2023)Dettmers, Pagnoni, Holtzman, and
  Zettlemoyer]{Dettmers:2023aa}
Tim Dettmers, Artidoro Pagnoni, Ari Holtzman, and Luke Zettlemoyer.
\newblock Qlora: Efficient finetuning of quantized llms, 2023.

\bibitem[Dodge et~al.(2021)Dodge, Sap, Marasovic, Agnew, Ilharco, Groeneveld,
  Mitchell, and Gardner]{Dodge:2021aa}
Jesse Dodge, Maarten Sap, Ana Marasovic, William Agnew, Gabriel Ilharco, Dirk
  Groeneveld, Margaret Mitchell, and Matt Gardner.
\newblock Documenting large webtext corpora: {A} case study on the colossal
  clean crawled corpus.
\newblock In Marie{-}Francine Moens, Xuanjing Huang, Lucia Specia, and
  Scott~Wen{-}tau Yih, editors, \emph{Proceedings of the 2021 Conference on
  Empirical Methods in Natural Language Processing, {EMNLP} 2021, Virtual Event
  / Punta Cana, Dominican Republic, 7-11 November, 2021}, pages 1286--1305.
  Association for Computational Linguistics, 2021.
\newblock \doi{10.18653/v1/2021.emnlp-main.98}.
\newblock URL \url{https://doi.org/10.18653/v1/2021.emnlp-main.98}.

\bibitem[Dong et~al.(2022)Dong, Xiao, Nozawa, Enomoto, and
  Oyamada]{Dong:2022aa}
Yuyang Dong, Chuan Xiao, Takuma Nozawa, Masafumi Enomoto, and Masafumi Oyamada.
\newblock Deepjoin: Joinable table discovery with pre-trained language models.
\newblock \emph{CoRR}, abs/2212.07588, 2022.
\newblock \doi{10.48550/arXiv.2212.07588}.
\newblock URL \url{https://doi.org/10.48550/arXiv.2212.07588}.

\bibitem[Dziri et~al.(2023)Dziri, Lu, Sclar, Li, Jiang, Lin, West, Bhagavatula,
  Bras, Hwang, Sanyal, Welleck, Ren, Ettinger, Harchaoui, and
  Choi]{Dziri:2023aa}
Nouha Dziri, Ximing Lu, Melanie Sclar, Xiang~Lorraine Li, Liwei Jiang,
  Bill~Yuchen Lin, Peter West, Chandra Bhagavatula, Ronan~Le Bras, Jena~D.
  Hwang, Soumya Sanyal, Sean Welleck, Xiang Ren, Allyson Ettinger, Za{\"{\i}}d
  Harchaoui, and Yejin Choi.
\newblock Faith and fate: Limits of transformers on compositionality.
\newblock \emph{CoRR}, abs/2305.18654, 2023.
\newblock \doi{10.48550/arXiv.2305.18654}.
\newblock URL \url{https://doi.org/10.48550/arXiv.2305.18654}.

\bibitem[Fan et~al.(2023)Fan, Wang, Li, and Miller]{Fan:2023aa}
Grace Fan, Jin Wang, Yuliang Li, and Ren{\'{e}}e~J. Miller.
\newblock Table discovery in data lakes: State-of-the-art and future
  directions.
\newblock In Sudipto Das, Ippokratis Pandis, K.~Sel{\c{c}}uk Candan, and Sihem
  Amer{-}Yahia, editors, \emph{Companion of the 2023 International Conference
  on Management of Data, {SIGMOD/PODS} 2023, Seattle, WA, USA, June 18-23,
  2023}, pages 69--75. {ACM}, 2023.
\newblock \doi{10.1145/3555041.3589409}.
\newblock URL \url{https://doi.org/10.1145/3555041.3589409}.

\bibitem[Foundation(2011)]{commoncrawl}
Common~Crawl Foundation.
\newblock Common crawl, 2011.
\newblock URL \url{https://commoncrawl.org}.

\bibitem[Gao et~al.(2021)Gao, Biderman, Black, Golding, Hoppe, Foster, Phang,
  He, Thite, Nabeshima, Presser, and Leahy]{DBLP:journals/corr/abs-2101-00027}
Leo Gao, Stella Biderman, Sid Black, Laurence Golding, Travis Hoppe, Charles
  Foster, Jason Phang, Horace He, Anish Thite, Noa Nabeshima, Shawn Presser,
  and Connor Leahy.
\newblock The pile: An 800gb dataset of diverse text for language modeling.
\newblock \emph{CoRR}, abs/2101.00027, 2021.
\newblock URL \url{https://arxiv.org/abs/2101.00027}.

\bibitem[Gehman et~al.(2020)Gehman, Gururangan, Sap, Choi, and
  Smith]{Gehman:2020aa}
Samuel Gehman, Suchin Gururangan, Maarten Sap, Yejin Choi, and Noah~A. Smith.
\newblock Realtoxicityprompts: Evaluating neural toxic degeneration in language
  models.
\newblock In Trevor Cohn, Yulan He, and Yang Liu, editors, \emph{Findings of
  the Association for Computational Linguistics: {EMNLP} 2020, Online Event,
  16-20 November 2020}, volume {EMNLP} 2020 of \emph{Findings of {ACL}}, pages
  3356--3369. Association for Computational Linguistics, 2020.
\newblock \doi{10.18653/v1/2020.findings-emnlp.301}.
\newblock URL \url{https://doi.org/10.18653/v1/2020.findings-emnlp.301}.

\bibitem[Green et~al.(2007)Green, Karvounarakis, and
  Tannen]{10.1145/1265530.1265535}
Todd~J. Green, Grigoris Karvounarakis, and Val Tannen.
\newblock Provenance semirings.
\newblock In \emph{Proceedings of the Twenty-Sixth ACM SIGMOD-SIGACT-SIGART
  Symposium on Principles of Database Systems}, PODS '07, pages 31--40, New
  York, NY, USA, 2007. Association for Computing Machinery.
\newblock ISBN 9781595936851.

\bibitem[Gualtieri and Yuhanna(2016)]{Gualtieri:2016aa}
Mike Gualtieri and Noel Yuhanna.
\newblock \emph{The Forrester Wave: Big Data Hadoop Distributions, Q1 2016}.
\newblock Forrester Research, Inc., January 2016.

\bibitem[Gudibande et~al.(2023)Gudibande, Wallace, Snell, Geng, Liu, Abbeel,
  Levine, and Song]{Gudibande:2023aa}
Arnav Gudibande, Eric Wallace, Charlie Snell, Xinyang Geng, Hao Liu, Pieter
  Abbeel, Sergey Levine, and Dawn Song.
\newblock The false promise of imitating proprietary llms, 2023.

\bibitem[Heaven(2022)]{Heaven:2022aa}
Will~Douglas Heaven.
\newblock Why meta's latest large language model survived only three days
  online.
\newblock \emph{MIT Technology Review}, November 2022.
\newblock URL
  \url{https://www.technologyreview.com/2022/11/18/1063487/meta-large-language-model-ai-only-survived-three-days-gpt-3-science/}.

\bibitem[Heidorn(2008)]{Heidorn:2008aa}
P.~Bryan Heidorn.
\newblock Shedding light on the dark data in the long tail of science.
\newblock \emph{Library trends}, 57\penalty0 (2):\penalty0 280--299, 2008.

\bibitem[Holtzman et~al.(2020)Holtzman, Buys, Du, Forbes, and
  Choi]{Holtzman:2020aa}
Ari Holtzman, Jan Buys, Li~Du, Maxwell Forbes, and Yejin Choi.
\newblock The curious case of neural text degeneration.
\newblock In \emph{8th International Conference on Learning Representations,
  {ICLR} 2020, Addis Ababa, Ethiopia, April 26-30, 2020}. OpenReview.net, 2020.
\newblock URL \url{https://openreview.net/forum?id=rygGQyrFvH}.

\bibitem[Hu et~al.(2019)Hu, Gaikwad, Bakker, Hulsebos, Zgraggen, Hidalgo,
  Kraska, Li, Satyanarayan, and Demiralp]{Hu:2019aa}
Kevin Hu, Neil Gaikwad, Michiel Bakker, Madelon Hulsebos, Emanuel Zgraggen,
  C\'{e}sar Hidalgo, Tim Kraska, Guoliang Li, Arvind Satyanarayan, and
  {\c{C}}a{\u{g}}atay Demiralp.
\newblock Viznet: {T}owards a large-scale visualization learning and
  benchmarking repository.
\newblock In \emph{Proceedings of the 2019 Conference on Human Factors in
  Computing Systems (CHI)}. ACM, 2019.

\bibitem[Huang et~al.(2022)Huang, Abbeel, Pathak, and Mordatch]{Huang:2022aa}
Wenlong Huang, Pieter Abbeel, Deepak Pathak, and Igor Mordatch.
\newblock Language models as zero-shot planners: Extracting actionable
  knowledge for embodied agents.
\newblock \emph{arXiv preprint arXiv:2201.07207}, 2022.

\bibitem[Hulsebos et~al.(2019)Hulsebos, Hu, Bakker, Zgraggen, Satyanarayan,
  Kraska, Demiralp, and Hidalgo]{DBLP:conf/kdd/HulsebosHBZSKDH19}
Madelon Hulsebos, Kevin~Zeng Hu, Michiel~A. Bakker, Emanuel Zgraggen, Arvind
  Satyanarayan, Tim Kraska, {\c{C}}agatay Demiralp, and C{\'{e}}sar~A. Hidalgo.
\newblock Sherlock: {A} deep learning approach to semantic data type detection.
\newblock In Ankur Teredesai, Vipin Kumar, Ying Li, R{\'{o}}mer Rosales,
  Evimaria Terzi, and George Karypis, editors, \emph{Proceedings of the 25th
  {ACM} {SIGKDD} International Conference on Knowledge Discovery {\&} Data
  Mining, {KDD} 2019, Anchorage, AK, USA, August 4-8, 2019}, pages 1500--1508.
  {ACM}, 2019.
\newblock \doi{10.1145/3292500.3330993}.
\newblock URL \url{https://doi.org/10.1145/3292500.3330993}.

\bibitem[Hulsebos et~al.(2021)Hulsebos, Demiralp, and
  Groth]{DBLP:journals/corr/abs-2106-07258}
Madelon Hulsebos, {\c{C}}agatay Demiralp, and Paul Groth.
\newblock Gittables: {A} large-scale corpus of relational tables.
\newblock \emph{CoRR}, abs/2106.07258, 2021.
\newblock URL \url{https://arxiv.org/abs/2106.07258}.

\bibitem[Iida et~al.(2021)Iida, Thai, Manjunatha, and Iyyer]{Iida:2021aa}
Hiroshi Iida, Dung Thai, Varun Manjunatha, and Mohit Iyyer.
\newblock {TABBIE:} pretrained representations of tabular data.
\newblock In Kristina Toutanova, Anna Rumshisky, Luke Zettlemoyer, Dilek
  Hakkani{-}T{\"{u}}r, Iz~Beltagy, Steven Bethard, Ryan Cotterell, Tanmoy
  Chakraborty, and Yichao Zhou, editors, \emph{Proceedings of the 2021
  Conference of the North American Chapter of the Association for Computational
  Linguistics: Human Language Technologies, {NAACL-HLT} 2021, Online, June
  6-11, 2021}, pages 3446--3456. Association for Computational Linguistics,
  2021.
\newblock \doi{10.18653/v1/2021.naacl-main.270}.
\newblock URL \url{https://doi.org/10.18653/v1/2021.naacl-main.270}.

\bibitem[Kandel et~al.(2011)Kandel, Paepcke, Hellerstein, and
  Heer]{DBLP:conf/chi/KandelPHH11}
Sean Kandel, Andreas Paepcke, Joseph~M. Hellerstein, and Jeffrey Heer.
\newblock Wrangler: interactive visual specification of data transformation
  scripts.
\newblock In Desney~S. Tan, Saleema Amershi, Bo~Begole, Wendy~A. Kellogg, and
  Manas Tungare, editors, \emph{Proceedings of the International Conference on
  Human Factors in Computing Systems, {CHI} 2011, Vancouver, BC, Canada, May
  7-12, 2011}, pages 3363--3372. {ACM}, 2011.
\newblock \doi{10.1145/1978942.1979444}.
\newblock URL \url{https://doi.org/10.1145/1978942.1979444}.

\bibitem[Kayali et~al.(2023)Kayali, Lykov, Fountalis, Vasiloglou, Olteanu, and
  Suciu]{Kayali:2023aa}
Moe Kayali, Anton Lykov, Ilias Fountalis, Nikolaos Vasiloglou, Dan Olteanu, and
  Dan Suciu.
\newblock {CHORUS:} foundation models for unified data discovery and
  exploration.
\newblock \emph{CoRR}, abs/2306.09610, 2023.
\newblock \doi{10.48550/arXiv.2306.09610}.
\newblock URL \url{https://doi.org/10.48550/arXiv.2306.09610}.

\bibitem[Khatiwada et~al.(2023)Khatiwada, Fan, Shraga, Chen, Gatterbauer,
  Miller, and Riedewald]{Khatiwada:2023aa}
Aamod Khatiwada, Grace Fan, Roee Shraga, Zixuan Chen, Wolfgang Gatterbauer,
  Ren{\'{e}}e~J. Miller, and Mirek Riedewald.
\newblock {SANTOS:} relationship-based semantic table union search.
\newblock \emph{Proc. {ACM} Manag. Data}, 1\penalty0 (1):\penalty0 9:1--9:25,
  2023.
\newblock \doi{10.1145/3588689}.
\newblock URL \url{https://doi.org/10.1145/3588689}.

\bibitem[Koleva et~al.(2021)Koleva, Ringsquandl, Joblin, and
  Tresp]{DBLP:journals/corr/abs-2110-15132}
Aneta Koleva, Martin Ringsquandl, Mitchell Joblin, and Volker Tresp.
\newblock Generating table vector representations.
\newblock \emph{CoRR}, abs/2110.15132, 2021.
\newblock URL \url{https://arxiv.org/abs/2110.15132}.

\bibitem[Kushmerick et~al.(1997)Kushmerick, Weld, and
  Doorenbos]{DBLP:conf/ijcai/KushmerickWD97}
Nicholas Kushmerick, Daniel~S. Weld, and Robert~B. Doorenbos.
\newblock Wrapper induction for information extraction.
\newblock In \emph{Proceedings of the Fifteenth International Joint Conference
  on Artificial Intelligence, {IJCAI} 97, Nagoya, Japan, August 23-29, 1997, 2
  Volumes}, pages 729--737. Morgan Kaufmann, 1997.

\bibitem[Langer et~al.(1978)Langer, Blank, and
  Chanowitz]{langer1978mindlessness}
Ellen~J Langer, Arthur Blank, and Benzion Chanowitz.
\newblock The mindlessness of ostensibly thoughtful action: The role of"
  placebic" information in interpersonal interaction.
\newblock \emph{Journal of personality and social psychology}, 36\penalty0
  (6):\penalty0 635, 1978.

\bibitem[Langford(1999)]{Langford:1999aa}
David Langford.
\newblock Comp.basilisk faq.
\newblock \emph{Nature}, 402\penalty0 (6761):\penalty0 465--465, 1999.
\newblock \doi{10.1038/44964}.
\newblock URL \url{https://doi.org/10.1038/44964}.

\bibitem[Levenshtein(1966)]{levenshtein1966binary}
Vladimir Levenshtein.
\newblock Binary codes capable of correcting deletions, insertions and
  reversals.
\newblock In \emph{Soviet Physics Doklady}, volume~10, page 707, 1966.

\bibitem[Lin et~al.(2022)Lin, Hilton, and Evans]{Lin:2022aa}
Stephanie Lin, Jacob Hilton, and Owain Evans.
\newblock {T}ruthful{QA}: Measuring how models mimic human falsehoods.
\newblock In \emph{Proceedings of the 60th Annual Meeting of the Association
  for Computational Linguistics (Volume 1: Long Papers)}, pages 3214--3252,
  Dublin, Ireland, May 2022. Association for Computational Linguistics.
\newblock \doi{10.18653/v1/2022.acl-long.229}.
\newblock URL \url{https://aclanthology.org/2022.acl-long.229}.

\bibitem[Mallen et~al.(2022)Mallen, Asai, Zhong, Das, Hajishirzi, and
  Khashabi]{Mallen:2022aa}
Alex Mallen, Akari Asai, Victor Zhong, Rajarshi Das, Hannaneh Hajishirzi, and
  Daniel Khashabi.
\newblock When not to trust language models: Investigating effectiveness and
  limitations of parametric and non-parametric memories, 2022.

\bibitem[Mendes et~al.(2012)Mendes, Jakob, and
  Bizer]{DBLP:conf/lrec/MendesJB12}
Pablo~N. Mendes, Max Jakob, and Christian Bizer.
\newblock Dbpedia: {A} multilingual cross-domain knowledge base.
\newblock In Nicoletta Calzolari, Khalid Choukri, Thierry Declerck, Mehmet~Ugur
  Dogan, Bente Maegaard, Joseph Mariani, Jan Odijk, and Stelios Piperidis,
  editors, \emph{Proceedings of the Eighth International Conference on Language
  Resources and Evaluation, {LREC} 2012, Istanbul, Turkey, May 23-25, 2012},
  pages 1813--1817. European Language Resources Association {(ELRA)}, 2012.
\newblock URL
  \url{http://www.lrec-conf.org/proceedings/lrec2012/summaries/570.html}.

\bibitem[Narayan et~al.(2022)Narayan, Chami, Orr, and R{\'{e}}]{Narayan:2022ab}
Avanika Narayan, Ines Chami, Laurel~J. Orr, and Christopher R{\'{e}}.
\newblock Can foundation models wrangle your data?
\newblock \emph{Proc. {VLDB} Endow.}, 16\penalty0 (4):\penalty0 738--746, 2022.
\newblock URL \url{https://www.vldb.org/pvldb/vol16/p738-narayan.pdf}.

\bibitem[Nargesian et~al.(2019)Nargesian, Zhu, Miller, Pu, and
  Arocena]{DBLP:journals/pvldb/NargesianZMPA19}
Fatemeh Nargesian, Erkang Zhu, Ren{\'{e}}e~J. Miller, Ken~Q. Pu, and
  Patricia~C. Arocena.
\newblock Data lake management: Challenges and opportunities.
\newblock \emph{Proc. {VLDB} Endow.}, 12\penalty0 (12):\penalty0 1986--1989,
  2019.
\newblock \doi{10.14778/3352063.3352116}.
\newblock URL \url{http://www.vldb.org/pvldb/vol12/p1986-nargesian.pdf}.

\bibitem[Nargesian et~al.(2023)Nargesian, Pu, Bashardoost, Zhu, and
  Miller]{DBLP:journals/tkde/NargesianPBZM23}
Fatemeh Nargesian, Ken~Q. Pu, Bahar~Ghadiri Bashardoost, Erkang Zhu, and
  Ren{\'{e}}e~J. Miller.
\newblock Data lake organization.
\newblock \emph{{IEEE} Trans. Knowl. Data Eng.}, 35\penalty0 (1):\penalty0
  237--250, 2023.
\newblock \doi{10.1109/TKDE.2021.3091101}.
\newblock URL \url{https://doi.org/10.1109/TKDE.2021.3091101}.

\bibitem[of~Licensing(2023)]{Licensing:2023aa}
Washington State~Department of~Licensing.
\newblock Electric vehicle population data electric vehicle population data, 04
  2023.
\newblock URL
  \url{https://catalog.data.gov/dataset/electric-vehicle-population-data}.

\bibitem[Olsson et~al.(2022)Olsson, Elhage, Nanda, Joseph, DasSarma, Henighan,
  Mann, Askell, Bai, Chen, Conerly, Drain, Ganguli, Hatfield{-}Dodds,
  Hernandez, Johnston, Jones, Kernion, Lovitt, Ndousse, Amodei, Brown, Clark,
  Kaplan, McCandlish, and Olah]{Olsson:2022aa}
Catherine Olsson, Nelson Elhage, Neel Nanda, Nicholas Joseph, Nova DasSarma,
  Tom Henighan, Ben Mann, Amanda Askell, Yuntao Bai, Anna Chen, Tom Conerly,
  Dawn Drain, Deep Ganguli, Zac Hatfield{-}Dodds, Danny Hernandez, Scott
  Johnston, Andy Jones, Jackson Kernion, Liane Lovitt, Kamal Ndousse, Dario
  Amodei, Tom Brown, Jack Clark, Jared Kaplan, Sam McCandlish, and Chris Olah.
\newblock In-context learning and induction heads.
\newblock \emph{CoRR}, abs/2209.11895, 2022.
\newblock \doi{10.48550/ARXIV.2209.11895}.
\newblock URL \url{https://doi.org/10.48550/arXiv.2209.11895}.

\bibitem[Ouyang et~al.(2022)Ouyang, Wu, Jiang, Almeida, Wainwright, Mishkin,
  Zhang, Agarwal, Slama, Ray, Schulman, Hilton, Kelton, Miller, Simens, Askell,
  Welinder, Christiano, Leike, and Lowe]{DBLP:journals/corr/abs-2203-02155}
Long Ouyang, Jeff Wu, Xu~Jiang, Diogo Almeida, Carroll~L. Wainwright, Pamela
  Mishkin, Chong Zhang, Sandhini Agarwal, Katarina Slama, Alex Ray, John
  Schulman, Jacob Hilton, Fraser Kelton, Luke Miller, Maddie Simens, Amanda
  Askell, Peter Welinder, Paul~F. Christiano, Jan Leike, and Ryan Lowe.
\newblock Training language models to follow instructions with human feedback.
\newblock \emph{CoRR}, abs/2203.02155, 2022.
\newblock \doi{10.48550/arXiv.2203.02155}.
\newblock URL \url{https://doi.org/10.48550/arXiv.2203.02155}.

\bibitem[Papenmeier et~al.(2021)Papenmeier, Kr{\"a}mer, Friedrich, Hienert, and
  Kern]{papenmeier2021genuine}
Andrea Papenmeier, Thomas Kr{\"a}mer, Tanja Friedrich, Daniel Hienert, and
  Dagmar Kern.
\newblock Genuine information needs of social scientists looking for data.
\newblock \emph{Proceedings of the Association for Information Science and
  Technology}, 58\penalty0 (1):\penalty0 292--302, 2021.

\bibitem[Rae et~al.(2021)Rae, Borgeaud, Cai, Millican, Hoffmann, Song,
  Aslanides, Henderson, Ring, Young, Rutherford, Hennigan, Menick, Cassirer,
  Powell, van~den Driessche, Hendricks, Rauh, Huang, Glaese, Welbl, Dathathri,
  Huang, Uesato, Mellor, Higgins, Creswell, McAleese, Wu, Elsen, Jayakumar,
  Buchatskaya, Budden, Sutherland, Simonyan, Paganini, Sifre, Martens, Li,
  Kuncoro, Nematzadeh, Gribovskaya, Donato, Lazaridou, Mensch, Lespiau,
  Tsimpoukelli, Grigorev, Fritz, Sottiaux, Pajarskas, Pohlen, Gong, Toyama,
  de~Masson~d'Autume, Li, Terzi, Mikulik, Babuschkin, Clark, de~Las~Casas, Guy,
  Jones, Bradbury, Johnson, Hechtman, Weidinger, Gabriel, Isaac, Lockhart,
  Osindero, Rimell, Dyer, Vinyals, Ayoub, Stanway, Bennett, Hassabis,
  Kavukcuoglu, and Irving]{Rae:2021aa}
Jack~W. Rae, Sebastian Borgeaud, Trevor Cai, Katie Millican, Jordan Hoffmann,
  H.~Francis Song, John Aslanides, Sarah Henderson, Roman Ring, Susannah Young,
  Eliza Rutherford, Tom Hennigan, Jacob Menick, Albin Cassirer, Richard Powell,
  George van~den Driessche, Lisa~Anne Hendricks, Maribeth Rauh, Po{-}Sen Huang,
  Amelia Glaese, Johannes Welbl, Sumanth Dathathri, Saffron Huang, Jonathan
  Uesato, John Mellor, Irina Higgins, Antonia Creswell, Nat McAleese, Amy Wu,
  Erich Elsen, Siddhant~M. Jayakumar, Elena Buchatskaya, David Budden, Esme
  Sutherland, Karen Simonyan, Michela Paganini, Laurent Sifre, Lena Martens,
  Xiang~Lorraine Li, Adhiguna Kuncoro, Aida Nematzadeh, Elena Gribovskaya,
  Domenic Donato, Angeliki Lazaridou, Arthur Mensch, Jean{-}Baptiste Lespiau,
  Maria Tsimpoukelli, Nikolai Grigorev, Doug Fritz, Thibault Sottiaux, Mantas
  Pajarskas, Toby Pohlen, Zhitao Gong, Daniel Toyama, Cyprien
  de~Masson~d'Autume, Yujia Li, Tayfun Terzi, Vladimir Mikulik, Igor
  Babuschkin, Aidan Clark, Diego de~Las~Casas, Aurelia Guy, Chris Jones, James
  Bradbury, Matthew~J. Johnson, Blake~A. Hechtman, Laura Weidinger, Iason
  Gabriel, William Isaac, Edward Lockhart, Simon Osindero, Laura Rimell, Chris
  Dyer, Oriol Vinyals, Kareem Ayoub, Jeff Stanway, Lorrayne Bennett, Demis
  Hassabis, Koray Kavukcuoglu, and Geoffrey Irving.
\newblock Scaling language models: Methods, analysis {\&} insights from
  training gopher.
\newblock \emph{CoRR}, abs/2112.11446, 2021.
\newblock URL \url{https://arxiv.org/abs/2112.11446}.

\bibitem[Ritze and Bizer(2017)]{DBLP:conf/edbt/RitzeB17}
Dominique Ritze and Christian Bizer.
\newblock Matching web tables to dbpedia - {A} feature utility study.
\newblock In Volker Markl, Salvatore Orlando, Bernhard Mitschang, Periklis
  Andritsos, Kai{-}Uwe Sattler, and Sebastian Bre{\ss}, editors,
  \emph{Proceedings of the 20th International Conference on Extending Database
  Technology, {EDBT} 2017, Venice, Italy, March 21-24, 2017}, pages 210--221.
  OpenProceedings.org, 2017.
\newblock \doi{10.5441/002/edbt.2017.20}.
\newblock URL \url{https://doi.org/10.5441/002/edbt.2017.20}.

\bibitem[Santos et~al.(2022)Santos, Bessa, Musco, and Freire]{Santos:2022aa}
A{\'{e}}cio S.~R. Santos, Aline Bessa, Christopher Musco, and Juliana Freire.
\newblock A sketch-based index for correlated dataset search.
\newblock In \emph{38th {IEEE} International Conference on Data Engineering,
  {ICDE} 2022, Kuala Lumpur, Malaysia, May 9-12, 2022}, pages 2928--2941.
  {IEEE}, 2022.
\newblock \doi{10.1109/ICDE53745.2022.00264}.
\newblock URL \url{https://doi.org/10.1109/ICDE53745.2022.00264}.

\bibitem[Shi et~al.(2023)Shi, Chen, Misra, Scales, Dohan, Chi, Sch{\"{a}}rli,
  and Zhou]{Shi:2023aa}
Freda Shi, Xinyun Chen, Kanishka Misra, Nathan Scales, David Dohan, Ed~H. Chi,
  Nathanael Sch{\"{a}}rli, and Denny Zhou.
\newblock Large language models can be easily distracted by irrelevant context.
\newblock \emph{CoRR}, abs/2302.00093, 2023.
\newblock \doi{10.48550/arXiv.2302.00093}.
\newblock URL \url{https://doi.org/10.48550/arXiv.2302.00093}.

\bibitem[Singhal et~al.(2022)Singhal, Azizi, Tu, Mahdavi, Wei, Chung, Scales,
  Tanwani, Cole{-}Lewis, Pfohl, Payne, Seneviratne, Gamble, Kelly,
  Sch{\"{a}}rli, Chowdhery, Mansfield, y~Arcas, Webster, Corrado, Matias, Chou,
  Gottweis, Tomasev, Liu, Rajkomar, Barral, Semturs, Karthikesalingam, and
  Natarajan]{Singhal:2022aa}
Karan Singhal, Shekoofeh Azizi, Tao Tu, S.~Sara Mahdavi, Jason Wei, Hyung~Won
  Chung, Nathan Scales, Ajay~Kumar Tanwani, Heather Cole{-}Lewis, Stephen
  Pfohl, Perry Payne, Martin Seneviratne, Paul Gamble, Chris Kelly, Nathaneal
  Sch{\"{a}}rli, Aakanksha Chowdhery, Philip~Andrew Mansfield,
  Blaise~Ag{\"{u}}era y~Arcas, Dale~R. Webster, Gregory~S. Corrado, Yossi
  Matias, Katherine Chou, Juraj Gottweis, Nenad Tomasev, Yun Liu, Alvin
  Rajkomar, Joelle~K. Barral, Christopher Semturs, Alan Karthikesalingam, and
  Vivek Natarajan.
\newblock Large language models encode clinical knowledge.
\newblock \emph{CoRR}, abs/2212.13138, 2022.
\newblock \doi{10.48550/arXiv.2212.13138}.
\newblock URL \url{https://doi.org/10.48550/arXiv.2212.13138}.

\bibitem[Suhara et~al.(2022)Suhara, Li, Li, Zhang, Demiralp, Chen, and
  Tan]{Suhara:2022aa}
Yoshihiko Suhara, Jinfeng Li, Yuliang Li, Dan Zhang, \c{C}a\u{g}atay Demiralp,
  Chen Chen, and Wang-Chiew Tan.
\newblock Annotating columns with pre-trained language models.
\newblock In \emph{Proceedings of the 2022 International Conference on
  Management of Data}. Association for Computing Machinery, 2022.
\newblock ISBN 9781450392495.
\newblock URL \url{https://doi.org/10.1145/3514221.3517906}.

\bibitem[Taylor et~al.(2022)Taylor, Kardas, Cucurull, Scialom, Hartshorn,
  Saravia, Poulton, Kerkez, and Stojnic]{Taylor:2022aa}
Ross Taylor, Marcin Kardas, Guillem Cucurull, Thomas Scialom, Anthony
  Hartshorn, Elvis Saravia, Andrew Poulton, Viktor Kerkez, and Robert Stojnic.
\newblock Galactica: {A} large language model for science.
\newblock \emph{CoRR}, abs/2211.09085, 2022.
\newblock \doi{10.48550/arXiv.2211.09085}.
\newblock URL \url{https://doi.org/10.48550/arXiv.2211.09085}.

\bibitem[Touvron et~al.(2023{\natexlab{a}})Touvron, Lavril, Izacard, Martinet,
  Lachaux, Lacroix, Rozi{\`{e}}re, Goyal, Hambro, Azhar, Rodriguez, Joulin,
  Grave, and Lample]{Touvron:2023aa}
Hugo Touvron, Thibaut Lavril, Gautier Izacard, Xavier Martinet, Marie{-}Anne
  Lachaux, Timoth{\'{e}}e Lacroix, Baptiste Rozi{\`{e}}re, Naman Goyal, Eric
  Hambro, Faisal Azhar, Aur{\'{e}}lien Rodriguez, Armand Joulin, Edouard Grave,
  and Guillaume Lample.
\newblock Llama: Open and efficient foundation language models.
\newblock \emph{CoRR}, abs/2302.13971, 2023{\natexlab{a}}.
\newblock \doi{10.48550/arXiv.2302.13971}.
\newblock URL \url{https://doi.org/10.48550/arXiv.2302.13971}.

\bibitem[Touvron et~al.(2023{\natexlab{b}})Touvron, Martin, Stone, Albert,
  Almahairi, Babaei, Bashlykov, Batra, Bhargava, Bhosale, Bikel, Blecher,
  Canton{-}Ferrer, Chen, Cucurull, Esiobu, Fernandes, Fu, Fu, Fuller, Gao,
  Goswami, Goyal, Hartshorn, Hosseini, Hou, Inan, Kardas, Kerkez, Khabsa,
  Kloumann, Korenev, Koura, Lachaux, Lavril, Lee, Liskovich, Lu, Mao, Martinet,
  Mihaylov, Mishra, Molybog, Nie, Poulton, Reizenstein, Rungta, Saladi,
  Schelten, Silva, Smith, Subramanian, Tan, Tang, Taylor, Williams, Kuan, Xu,
  Yan, Zarov, Zhang, Fan, Kambadur, Narang, Rodriguez, Stojnic, Edunov, and
  Scialom]{Touvron:2023ab}
Hugo Touvron, Louis Martin, Kevin Stone, Peter Albert, Amjad Almahairi, Yasmine
  Babaei, Nikolay Bashlykov, Soumya Batra, Prajjwal Bhargava, Shruti Bhosale,
  Dan Bikel, Lukas Blecher, Cristian Canton{-}Ferrer, Moya Chen, Guillem
  Cucurull, David Esiobu, Jude Fernandes, Jeremy Fu, Wenyin Fu, Brian Fuller,
  Cynthia Gao, Vedanuj Goswami, Naman Goyal, Anthony Hartshorn, Saghar
  Hosseini, Rui Hou, Hakan Inan, Marcin Kardas, Viktor Kerkez, Madian Khabsa,
  Isabel Kloumann, Artem Korenev, Punit~Singh Koura, Marie{-}Anne Lachaux,
  Thibaut Lavril, Jenya Lee, Diana Liskovich, Yinghai Lu, Yuning Mao, Xavier
  Martinet, Todor Mihaylov, Pushkar Mishra, Igor Molybog, Yixin Nie, Andrew
  Poulton, Jeremy Reizenstein, Rashi Rungta, Kalyan Saladi, Alan Schelten, Ruan
  Silva, Eric~Michael Smith, Ranjan Subramanian, Xiaoqing~Ellen Tan, Binh Tang,
  Ross Taylor, Adina Williams, Jian~Xiang Kuan, Puxin Xu, Zheng Yan, Iliyan
  Zarov, Yuchen Zhang, Angela Fan, Melanie Kambadur, Sharan Narang,
  Aur{\'{e}}lien Rodriguez, Robert Stojnic, Sergey Edunov, and Thomas Scialom.
\newblock Llama 2: Open foundation and fine-tuned chat models.
\newblock \emph{CoRR}, abs/2307.09288, 2023{\natexlab{b}}.
\newblock \doi{10.48550/arXiv.2307.09288}.
\newblock URL \url{https://doi.org/10.48550/arXiv.2307.09288}.

\bibitem[Trifacta(2023)]{trifacta_wrangler}
Trifacta.
\newblock Trifacta wrangler.
\newblock \url{https://cloud.trifacta.com}, 2023.
\newblock Accessed: 2023-04-10.

\bibitem[Trummer(2021)]{Trummer:2021aa}
Immanuel Trummer.
\newblock Can deep neural networks predict data correlations from column names?
\newblock \emph{CoRR}, abs/2107.04553, 2021.
\newblock URL \url{https://arxiv.org/abs/2107.04553}.

\bibitem[Trummer(2022)]{Trummer:2022aa}
Immanuel Trummer.
\newblock Towards nlp-enhanced data profiling tools.
\newblock In \emph{12th Conference on Innovative Data Systems Research, {CIDR}
  2022, Chaminade, CA, USA, January 9-12, 2022}. www.cidrdb.org, 2022.
\newblock URL \url{https://www.cidrdb.org/cidr2022/papers/a55-trummer.pdf}.

\bibitem[Vemprala et~al.(2023)Vemprala, Bonatti, Bucker, and
  Kapoor]{Vemprala:2023aa}
Sai Vemprala, Rogerio Bonatti, Arthur Bucker, and Ashish Kapoor.
\newblock Chatgpt for robotics: Design principles and model abilities.
\newblock Technical Report MSR-TR-2023-8, Microsoft, February 2023.
\newblock URL
  \url{https://www.microsoft.com/en-us/research/publication/chatgpt-for-robotics-design-principles-and-model-abilities/}.

\bibitem[Wei et~al.(2022{\natexlab{a}})Wei, Tay, Bommasani, Raffel, Zoph,
  Borgeaud, Yogatama, Bosma, Zhou, Metzler, Chi, Hashimoto, Vinyals, Liang,
  Dean, and Fedus]{Wei:2022aa}
Jason Wei, Yi~Tay, Rishi Bommasani, Colin Raffel, Barret Zoph, Sebastian
  Borgeaud, Dani Yogatama, Maarten Bosma, Denny Zhou, Donald Metzler, Ed~H.
  Chi, Tatsunori Hashimoto, Oriol Vinyals, Percy Liang, Jeff Dean, and William
  Fedus.
\newblock Emergent abilities of large language models, 2022{\natexlab{a}}.

\bibitem[Wei et~al.(2022{\natexlab{b}})Wei, Tay, Bommasani, Raffel, Zoph,
  Borgeaud, Yogatama, Bosma, Zhou, Metzler, Chi, Hashimoto, Vinyals, Liang,
  Dean, and Fedus]{Wei:2022ab}
Jason Wei, Yi~Tay, Rishi Bommasani, Colin Raffel, Barret Zoph, Sebastian
  Borgeaud, Dani Yogatama, Maarten Bosma, Denny Zhou, Donald Metzler, Ed~H.
  Chi, Tatsunori Hashimoto, Oriol Vinyals, Percy Liang, Jeff Dean, and William
  Fedus.
\newblock Emergent abilities of large language models.
\newblock \emph{Trans. Mach. Learn. Res.}, 2022, 2022{\natexlab{b}}.
\newblock URL \url{https://openreview.net/forum?id=yzkSU5zdwD}.

\bibitem[West et~al.(2022)West, Bhagavatula, Hessel, Hwang, Jiang, Bras, Lu,
  Welleck, and Choi]{West:2022aa}
Peter West, Chandra Bhagavatula, Jack Hessel, Jena~D. Hwang, Liwei Jiang,
  Ronan~Le Bras, Ximing Lu, Sean Welleck, and Yejin Choi.
\newblock Symbolic knowledge distillation: from general language models to
  commonsense models.
\newblock In Marine Carpuat, Marie{-}Catherine de~Marneffe, and Iv{\'{a}}n
  Vladimir~Meza Ru{\'{\i}}z, editors, \emph{Proceedings of the 2022 Conference
  of the North American Chapter of the Association for Computational
  Linguistics: Human Language Technologies, {NAACL} 2022, Seattle, WA, United
  States, July 10-15, 2022}, pages 4602--4625. Association for Computational
  Linguistics, 2022.
\newblock \doi{10.18653/v1/2022.naacl-main.341}.
\newblock URL \url{https://doi.org/10.18653/v1/2022.naacl-main.341}.

\bibitem[Yan and He(2020)]{DBLP:conf/sigmod/YanH20}
Cong Yan and Yeye He.
\newblock Auto-suggest: Learning-to-recommend data preparation steps using data
  science notebooks.
\newblock In David Maier, Rachel Pottinger, AnHai Doan, Wang{-}Chiew Tan,
  Abdussalam Alawini, and Hung~Q. Ngo, editors, \emph{Proceedings of the 2020
  International Conference on Management of Data, {SIGMOD} Conference 2020,
  online conference [Portland, OR, USA], June 14-19, 2020}, pages 1539--1554.
  {ACM}, 2020.
\newblock \doi{10.1145/3318464.3389738}.
\newblock URL \url{https://doi.org/10.1145/3318464.3389738}.

\bibitem[Yin et~al.(2020)Yin, Neubig, Yih, and Riedel]{DBLP:conf/acl/YinNYR20}
Pengcheng Yin, Graham Neubig, Wen{-}tau Yih, and Sebastian Riedel.
\newblock Tabert: Pretraining for joint understanding of textual and tabular
  data.
\newblock In Dan Jurafsky, Joyce Chai, Natalie Schluter, and Joel~R. Tetreault,
  editors, \emph{Proceedings of the 58th Annual Meeting of the Association for
  Computational Linguistics, {ACL} 2020, Online, July 5-10, 2020}, pages
  8413--8426. Association for Computational Linguistics, 2020.
\newblock \doi{10.18653/v1/2020.acl-main.745}.
\newblock URL \url{https://doi.org/10.18653/v1/2020.acl-main.745}.

\bibitem[Zhang et~al.(2016)Zhang, Shin, R{\'{e}}, Cafarella, and
  Niu]{Zhang:2016aa}
Ce~Zhang, Jaeho Shin, Christopher R{\'{e}}, Michael~J. Cafarella, and Feng Niu.
\newblock Extracting databases from dark data with deepdive.
\newblock In Fatma {\"{O}}zcan, Georgia Koutrika, and Sam Madden, editors,
  \emph{Proceedings of the 2016 International Conference on Management of Data,
  {SIGMOD} Conference 2016, San Francisco, CA, USA, June 26 - July 01, 2016},
  pages 847--859. {ACM}, 2016.
\newblock \doi{10.1145/2882903.2904442}.
\newblock URL \url{https://doi.org/10.1145/2882903.2904442}.

\bibitem[Zhang et~al.(2020)Zhang, Suhara, Li, Hulsebos, Demiralp, and
  Tan]{Zhang:2020aa}
Dan Zhang, Yoshihiko Suhara, Jinfeng Li, Madelon Hulsebos, {\c{C}}agatay
  Demiralp, and Wang{-}Chiew Tan.
\newblock Sato: Contextual semantic type detection in tables.
\newblock \emph{Proc. {VLDB} Endow.}, 13\penalty0 (11):\penalty0 1835--1848,
  2020.
\newblock URL \url{http://www.vldb.org/pvldb/vol13/p1835-zhang.pdf}.

\bibitem[Zhang et~al.(2023)Zhang, Press, Merrill, Liu, and Smith]{Zhang:2023aa}
Muru Zhang, Ofir Press, William Merrill, Alisa Liu, and Noah~A. Smith.
\newblock How language model hallucinations can snowball, 2023.

\bibitem[Zhao et~al.(2021)Zhao, Wallace, Feng, Klein, and Singh]{Zhao:2021aa}
Zihao Zhao, Eric Wallace, Shi Feng, Dan Klein, and Sameer Singh.
\newblock Calibrate before use: Improving few-shot performance of language
  models.
\newblock In Marina Meila and Tong Zhang, editors, \emph{Proceedings of the
  38th International Conference on Machine Learning, {ICML} 2021, 18-24 July
  2021, Virtual Event}, volume 139 of \emph{Proceedings of Machine Learning
  Research}, pages 12697--12706. {PMLR}, 2021.
\newblock URL \url{http://proceedings.mlr.press/v139/zhao21c.html}.

\bibitem[Zheng et~al.(2023)Zheng, Chiang, Sheng, Zhuang, Wu, Zhuang, Lin, Li,
  Li, Xing, Zhang, Gonzalez, and Stoica]{Zheng:2023aa}
Lianmin Zheng, Wei{-}Lin Chiang, Ying Sheng, Siyuan Zhuang, Zhanghao Wu,
  Yonghao Zhuang, Zi~Lin, Zhuohan Li, Dacheng Li, Eric~P. Xing, Hao Zhang,
  Joseph~E. Gonzalez, and Ion Stoica.
\newblock Judging llm-as-a-judge with mt-bench and chatbot arena.
\newblock \emph{CoRR}, abs/2306.05685, 2023.
\newblock \doi{10.48550/arXiv.2306.05685}.
\newblock URL \url{https://doi.org/10.48550/arXiv.2306.05685}.

\bibitem[Zhu et~al.(2019)Zhu, Deng, Nargesian, and Miller]{Zhu:2019aa}
Erkang Zhu, Dong Deng, Fatemeh Nargesian, and Ren{\'{e}}e~J. Miller.
\newblock {JOSIE:} overlap set similarity search for finding joinable tables in
  data lakes.
\newblock In Peter~A. Boncz, Stefan Manegold, Anastasia Ailamaki, Amol
  Deshpande, and Tim Kraska, editors, \emph{Proceedings of the 2019
  International Conference on Management of Data, {SIGMOD} Conference 2019,
  Amsterdam, The Netherlands, June 30 - July 5, 2019}, pages 847--864. {ACM},
  2019.
\newblock \doi{10.1145/3299869.3300065}.
\newblock URL \url{https://doi.org/10.1145/3299869.3300065}.

\end{thebibliography}
\bibliographystyle{plainnat}

\end{document}